\documentclass[12pt]{article}

\usepackage{amsmath,amssymb}
\usepackage{epsfig}
\usepackage{graphicx}
\usepackage{enumerate}

\newtheorem{theorem}{Theorem}[section]

\newtheorem{lemma}[theorem]{Lemma}

\newtheorem{example}[theorem]{Example}
\newtheorem{remark}[theorem]{Remark}

\newcommand{\e}{\mathrm{e}}
\newcommand{\eps}{\varepsilon}
\renewcommand{\i}{\mathrm{i}}

\renewcommand{\d}{\mathrm{d}}
\newcommand{\ds}{\displaystyle}

\newcommand{\dsfrac}{\ds\frac}

\newcommand{\N}{\mathbb{N}}
\newcommand{\OO}{\mathcal{O}}
\newcommand{\R}{\mathbb{R}}
\newcommand{\Z}{\mathbb{Z}}
\renewcommand{\(}{\left(}
\renewcommand{\)}{\right)}

\begin{document}

\noindent\textbf{\Large Curvature-induced bound states in Robin \\[.2em] waveguides and their asymptotical properties}

\begin{quote}
{\large Pavel Exner$^{1,2}$ and Alexander Minakov$^{1,3}$}
\\{\emph{\small $^1$Doppler Institute for Mathematical Physics and Applied Mathematics,
Czech Technical University in Prague, B\v{r}ehov\'a 7, 11519 Prague, Czech Republic}
\\{\em $^2$Department of Theoretical Physics, Nuclear Physics Institute ASCR, Hlavní 130,
25068 \v{R}e\v{z} near Prague, Czech Republic}
\\{\em $^3$Department of Physics, Faculty of Nuclear
Science and Physical Engineering, Czech Technical University in
Prague, Pohrani\v{c}ní 1288/1, 40501 D\v{e}\v{c}ín, Czech  Republic}
\\{exner@ujf.cas.cz, minakov.ilt@gmail.com}}
\end{quote}

\begin{quote}
We analyze bound states of Robin Laplacian in infinite planar domains with a smooth boundary, in particular, their relations to the geometry of the latter. The domains considered have locally straight boundary being, for instance, locally deformed halfplanes or wedges, or infinite strips, alternatively they are the exterior of a bounded obstacle. In the situation when the Robin condition is strongly attractive, we derive a two-term asymptotic formula in which the next-to-leading term is determined by the extremum of the boundary curvature. We also discuss the non-asymptotic case of attractive boundary interaction and show that the discrete spectrum is nonempty if the domain is a local deformation of a halfplane or a wedge of angle less than $\pi$, and it is void if the domain is concave.
\end{quote}

%\begin{classification} 35P15, 35J05. \end{classification}

%\begin{keywords} Laplacian, Robin problem, eigenvalue asymptotics. \end{keywords}

%%%%%%%%%%%%%%%%%%%%%%%%%%%%%%%%%%%%%%%%%%%%%%%%%%%%%%%
\section{Introduction} \label{sect: intro}

The task of control motion of quantum particles guiding them in a desired direction has both the theoretical and practical significance. Most often this problem is addressed in the situation when the motion is confined to a strip or a tube which models real-world objects such as semiconductor quantum wires, carbon nanotubes, etc. The boundary condition involved are at that typically Dirichlet, modeling a hard-wall boundary, or Neumann. A single boundary cannot produce in this situation a guided motion unless an external field is added \cite{FGW00,MMP99}.

The situation changes if the boundary is described by a mixed-type condition as in \eqref{problem_0} below, conventionally called Robin, representing and attractive contact interaction, in other words, with the parameter $\beta>0$. If a two-dimensional particle is confined to a halfplane with such a boundary, its spectrum is purely absolutely continuous covering the interval $[-\beta^2,\infty)$ and contains a component describing states moving along the boundary, in particular, any states referring to the spectral projection on the interval $[-\beta^2,0)$ have this property. The spectral simple picture becomes more complicated, however, when the boundary is geometrically nontrivial and the corresponding Laplacian may have a nonempty discrete spectrum. A discussion of such eigenvalues is the main topic of this paper.

Our main result concerns the asymptotic behavior of the eigenvalue in the strong coupling case, $\beta\to\infty$. We are going to consider planar domains the boundary of which is an infinite smooth curve without self-intersections assuming that its curvature decays fast enough outside a compact. The method we shall use combines a bracketing argument with spectral analysis of Laplacians in a curved strip. It was first proposed in Ref.~\cite{Exner_Yoshitomi 2002} dealing with curve-supported $\delta$ interactions in the plane. Its `one-sided' version suitable for application to domains with a boundary proved to be different, however, due to the presence of an extra term which changes substantially the result \cite{EMP} leading to asymptotic inequalities only. A two-term asymptotic expansion can be nevertheless obtain if one combines this conclusion with a variational estimate devised by K.~Pankrashkin \cite{Pankrashkin 2013} for treatment of the ground state. We note that the smoothness of the boundary is essential, once it is allowed to have angles the asymptotics changes in the leading order \cite{Levitin_Parnovskii 2008}.

In contrast to Ref.~\cite{EMP} where Robin billiards were discussed we consider here infinite domains where the essential spectrum is nonempty and the discrete one may not exist. The asymptotic formula we are going to derive will yield sufficient conditions for its existence and nonexistence in the strong coupling regime. In addition, one is able to show that if such geometrically induced bound states exist, a sufficiently strong boundary attraction can produce any prescribed finite number of them. Furthermore, with small modifications the technique will allow us to derive analogous asymptotic expansion for Robin waveguides where the existence of the discrete spectrum is known \cite{Jilek_06}, as well for domains being the exterior of a compact obstacle.

The result concerning a single infinite boundary curve raises naturally the question about existence of bound states beyond the asymptotic regime. For simplicity we shall suppose that the parts of the boundary outside a compact are straight and not parallel, on the other hand we weaken the requirement on the local smoothness. The essential spectrum then coincides with the interval $[-\beta^2,\infty)$ and using variational arguments we are going to show that the discrete spectrum is nonempty provided the `internal' angle between the two boundary asymptotes does not exceed $\pi$, except for the trivial case when the domain in question is a halfplane. On the other hand, while bound states may sometimes exist even for the asymptote angle in $(\pi,2\pi)$, the discrete spectrum is empty if the domain in question is concave.

The paper is organized as follows. In the next section we formulate the state the problem properly and formulate the results about the strong-coupling asymptotic behavior of the discrete spectrum for the case of a single infinite boundary. They will be subsequently proved in Section~\ref{sect: Proof half plane}; some parts of the proofs follows closely the analogous arguments in Ref.~\cite{EMP} and will be presented only briefly. Consequences for a curved Robin waveguide are indicated in Section~\ref{sect: waveguide}. Then we pass to the non-asymptotic regime and prove in Section~\ref{sect: Proof of existence of bound state} the above mentioned results about the spectrum. Finally, in Section~\ref{sect: infinite domain} we shall present the asymptotic result for the Robin problem in the exterior of a bounded domain.

%%%%%%%%%%%%%%%%%%%%%%%%%%%%%%%%%%%%%%%%%%%%%%%%%%%%%%%
\section{Robin problem in a domain with single infinite boundary}
\label{sect: half-plane}

Let us now state our Robin problem properly. Consider a domain $\Omega\subset\R^2$ the boundary of which is an infinite smooth curve $\Gamma$ without self-intersections, and assume that
 % -------------- %
\begin{enumerate}[(i)]
\setlength{\itemsep}{-3pt}
 % -------------- %
\item there is an $a>0$ such that any two points of $\Gamma$ the arc-length distance of which is larger that $3a$ have disjoint $a$-neighborhoods in $\Omega$, or equivalently, there is a $a$-neighborhood of the boundary in $\Omega$ which does not intersect itself,
 % -------------- %
\item $\Gamma:\:\R\to\R^2$ is a $C^4$-smooth curve, without loss of generality we may parametrize itself by its arc length, $\Gamma(s) = (\Gamma_1,\Gamma_2)$. The orientation is chosen in such a way that $\Omega$ lies to the left of $\Gamma$ if we follow it in the direction of increasing $s$,
 % -------------- %
\item the signed curvature $\gamma(s)= \Gamma_1'(s)\Gamma_2''(s)-\Gamma_2'(s)\Gamma_1''(s)$ of $\Gamma$ satisfies the bound $|\gamma(s)| \le c \langle s\rangle^{-1-\eps}$ for some numbers $c,\eps>0$, where as usual we put $\langle s\rangle := \sqrt{1+s^2}$.
 % -------------- %
\item the first two derivatives of $\gamma$ are bounded, $\gamma_+':=\max\limits_{s\in\R}|\gamma'(s)|<\infty$ and $\gamma_+'':=\max\limits_{s\in\R}|\gamma''(s)|<\infty$.
 % -------------- %
\end{enumerate}
 % -------------- %
For future purposes we introduce the following quantities
 % -------------- %
\begin{equation} \label{curv_extreme}
\gamma_+:=\max\limits_{s\in\R}|\gamma(s)|<\infty\,,\quad \gamma^*:=\max\limits_{s\in\R}\gamma(s)\,,\quad \gamma_*:=\min\limits_{s\in\R}\gamma(s)\,,
\end{equation}
 % -------------- %
which are finite due to the assumptions.

 % -------------- %
\begin{remark} \label{rem: spiral}
{\rm The assumption (iii) guarantees that the tangent vector of $\Gamma$ has limits as $|s|\to\infty$. On the other hand, it does not require the existence of asymptotes, as an example one can take a domain delineated by a parabola, where $\gamma(s) = \OO(|s|^{-3/2})$. It will be clear from the discussion in the next section, however, that the method works even for curves with a slower curvature decay allowing thus for more `exotic' domains, for instance, an $\Omega$ which is outside a compact (the interior of which contains the origin of coordinates) bordered by a pair of logarithmic spirals given by parametric equations $r=a_j\, \e^{b\theta}$ with $a_1<a_2<a_1\, \e^{2\pi b}$.}
\end{remark}
 % -------------- %

Let us now consider the following boundary-value problem in $\Omega$:
 % -------------- %
\begin{eqnarray}
-\Delta f=\lambda f\; &\textrm {in}& \Omega\,, \nonumber \\
[-.5em] && \label{problem_0} \\ [-.5em] \frac{\partial f}{\partial
n}=\beta f\; &\textrm{on}& \partial\Omega=\Gamma \;\:\textrm{with}\; \beta>0\,,\nonumber
\end{eqnarray}
 % -------------- %
where the symbol $\frac{\partial}{\partial n}$ denotes the outside normal derivative. It is straightforward to check that the quadratic form
 % -------------- %
\begin{equation}\label{differential form}
q_{\beta}[f] =\|\nabla
f\|^2_{L^2(\Omega)}-\beta\int\limits_{\Gamma}|f(x)|^2\d s
\end{equation}
 % -------------- %
with $\mathrm{Dom}(q_{\beta}) = H^{1}(\Omega)$ is closed and below bounded; we denote by $H_{\beta}$ the unique self-adjoint operator associated with it. In general $H_{\beta}$ may not have a nonempty discrete spectrum --- think of the situation when $\Omega$ is halfplane --- but we shall assume that $\sigma_\mathrm{disc}(H_\beta)\ne\emptyset$ and denote by $\lambda_j$ its $j$-th eigenvalue. The existence of such eigenvalues will be a matter of a further discussion. Our main goal is to analyze the asymptotic behavior of $\lambda_j= \lambda_j(\beta)$ as the parameter $\beta$ tends to infinity. To state the results we have to introduce the operator
 % -------------- %
\begin{equation}\label{longitudinal_operator_S}
S=-\dsfrac{\d^2}{\d s^2}-\frac{1}{4}\gamma^2(s)
\quad\textrm{ in } L^2(\R)
\end{equation}
 % -------------- %
with the domain $H^2(\R)$. In view of assumption (iii) its essential spectrum is $(0,\infty)$ and the discrete spectrum is nonempty unless $\gamma=0$, under the decay hypothesis we made it is finite. We denote by $\mu_j$ the $j$-th eigenvalue of $S_0$, provided it exists. Our first main result then reads as follows.

 % -------------- %
\begin{theorem} \label{main theorem_half-plane}
Adopt the assumptions (i)--(iv) and denote $\#\sigma_\mathrm{disc}(S)=M$, the multiplicity taken into account. If $\gamma^*>0$, then to any $N\in\N$ there is a $\beta_N>0$ such that the problem (\ref{problem_0}) has at least $N$ eigenvalues, again counted with their multiplicity, for all $\beta>\beta_N$, satisfying the he following asymptotic expansion
 % -------------- %
$$
\lambda_j(\beta) = -\beta^2-\gamma^*\beta+\OO\big(\beta^{2/3}\big)\,.
$$
 % -------------- %
The lower bound can be made more precise, specifically
 % -------------- %
\begin{eqnarray}\label{estimate_lambda n}
\lambda_j(\beta) &\!\!\ge\!\!& -\(\beta+\frac{\gamma^*}{2}\)^2+\mu_j+
\OO\(\frac{\log\beta}{\beta}\)\,, \quad
1\le j\le M\,,
\\
\lambda_j(\beta) &\!\!\ge\!\!& -\(\beta+\frac{\gamma^*}{2}\)^2+
\OO\(\frac{\log\beta}{\beta}\)\,, \quad j>M\,.
\label{estimate_lambda n_}
\end{eqnarray}
 % -------------- %
\end{theorem}
 % -------------- %

\noindent We see that in the asymptotic regime the condition $\gamma^*>0$, in other words, a local convexity of the boundary, produces a discrete spectrum of any finite cardinality. Note that this includes situations when the boundary is a local deformation of a straight line, since in such a case we have $\int_\R \gamma(s)\, \d s=0$, and consequently, the curvature has to be sign changing.

 % -------------- %
\begin{remark} \label{rem: quadrant}
{\rm Let us stress that the claim about the existence of any finite number of bound states for $\beta$ large enough requires a sufficient smoothness of the boundary. Not only the proof presented in Sec.~\ref{sect: Proof half plane} needs this assumption, but the result is not valid without it in general. As an example consider $\Omega$ in the form of a quadrant, $\{(x,y)\in\R^2:\: x>0,\: y>0\}$ with the boundary consisting of two halflines meeting at the right angle. The spectrum is easily found by separation of variables: we have $\sigma_\mathrm{ess}(H_\beta) = [-\beta^2,\infty)$ and for any $\beta>0$ there is exactly one simple eigenvalue equal to $-2\beta^2$.}
\end{remark}
 % -------------- %

%%%%%%%%%%%%%%%%%%%%%%%%%%%%%%%%%%%%%%%%%%%%%%%%%%%%%%%
\section{Proof of Theorem \ref{main theorem_half-plane}}
\label{sect: Proof half plane}

Let us start with introducing quadratic forms and operators which we shall need in the argument. To this aim, we need the following result analogous to Lemma~2.1 of Ref.~\cite{Exner_Yoshitomi 2002}.

 % -------------- %
\begin{lemma}\label{lemma_existence of C_u}
There is an $a_1>0$ such that the map $\Phi$,
 % -------------- %
\[
\R\times(0,a)\ni(s,u)\mapsto(\Gamma_1(s)-u\Gamma_2'(s),\Gamma_2(s)+u\Gamma_1'(s))\in\mathbb{R}^2\,,
\]
 % -------------- %
is injective for any $a\in(0,a_1].$
\end{lemma}
 % -------------- %

\medskip

\noindent We skip the proof which is basically the same as in Ref.~\cite{Exner_Yoshitomi 2002} in checking that $\Phi$ is locally a diffeomorphism. The only new element here is the non-compactness of $\Phi(\R\times(0,a))$ which requires the assumption (i) above to ensure the injectivity.

Choose now an $a\in(0,a_1]$ to be specified later and denote by $\Sigma_{a}$ the strip neighborhood of $\Gamma\equiv\Gamma^0$ of the width $a$, in other words
 % -------------- %
\[
\Sigma_{a}:=\Phi(\R\times(0,a))\,.
\]
 % -------------- %
Then $\Omega\setminus\overline\Sigma_a=:\Lambda_a$ is an unbounded simply connected domain with the boundary which we denote as $\Gamma^a$. We define the quadratic forms
 % -------------- %
\begin{eqnarray*}
&& q_{a,\beta}^{D}[f]:=\|\nabla
f\|^2_{\Sigma_a}-\beta\int\limits_{\Gamma^0}|f(x)|^2\d s\quad\textrm{
for }\: f\in\left\{f\in H^1(\Sigma_a): f|_{\Gamma^a}=0\right\}, \\
&& q_{a,\beta}^{N}[f]:=\|\nabla
f\|^2_{\Sigma_a}-\beta\int\limits_{\Gamma^0}|f(x)|^2\d s\quad\textrm{
for }\: f\in H^1(\Sigma_a),
\end{eqnarray*}
 % -------------- %
which are easily checked to be closed and bounded from below, and denote by $L_{a,\beta}^{D}$, $L_{a,\beta}^{N}$ the unique  self-adjoint operators associated with $q_{a,\beta}^{D}$, $q_{a,\beta}^{N}$, respectively. The first key component of the proof is a Dirichlet-Neumann bracketing --- see  \cite[Sec.~XIII.15, Prop.~4]{Reed Simon 4 1978} --- in our case it consists of imposing  additional boundary condition at $\Gamma^a$. This yields the bounds
 % -------------- %
\begin{equation}\label{inequalities H_beta}
L_{a,\beta}^{N}\oplus(-\Delta^N_{\Lambda_a})\leq H_{\beta,0}\leq
L_{a,\beta}^{D}\oplus (-\Delta^D_{\Lambda_a})
\end{equation}
 % -------------- %
in $L^2(\Omega)= L^2(\Sigma_a)\oplus L^2(\Lambda_a).$ Since the estimating operators have a direct-sum structure and the  terms in the inequalities (\ref{inequalities H_beta}) referring to $\Lambda_a$ are positive, in order to estimate the (negative) eigenvalues of $H_{\beta,0}$ it is sufficient to estimate those of $L^{D}_{a,\beta}$ and $L^{N}_{a,\beta}.$

To achieve this, we bring in the second main ingredient of the proof introducing a `straightening' transformation in the spirit of Ref.~\cite{Exner_Seba 1989} which produces a pair of operators in $L^2(\mathbb{R}\times(0,a))$ that are unitarily equivalent to $L_{a,\beta}^{D}$ and $L_{a,\beta}^{N},$ respectively. This is achieved by introducing the following change of variables,
 % -------------- %
\begin{equation} \label{eff-wf}
f(x_1,x_2)=\frac{1}{(1-u\gamma(s))^{1/2}}\ \varphi(s,u)\,;
\end{equation}
 % -------------- %
then it is straightforward to check that for any function $f\in H^2(\Sigma_a)$ we have also $\varphi\in H^2(\R\times(0,a))$ and
 % -------------- %
\begin{eqnarray*}
\lefteqn{(|f_{x_1}|^2+|f_{x_2}|^2)(x_1,x_2)=\left[\frac{1}{(1-u\gamma)^2}\left|\frac{\partial\varphi}{\partial s}\right|^2+
\left|\frac{\partial\varphi}{\partial u}\right|^2 +\widetilde
V|\varphi|^2\right.} \\ && \left.+\frac{u\gamma'}{2(1-u\gamma)^3}\(\varphi\
\overline{\frac{\partial\varphi} {\partial s}}+\overline{\varphi}\
\frac{\partial\varphi} {\partial
s}\)+\frac{\gamma}{2(1-u\gamma)}\(\varphi\
\overline{\frac{\partial\varphi} {\partial u}}+\overline{\varphi}\
\frac{\partial\varphi} {\partial u}\)\right](s,u)
\end{eqnarray*}
 % -------------- %
with
 % -------------- %
\[
\widetilde
V(s,u):=\frac{\gamma^2(s)}{4(1-u\gamma(s))^2}+\frac{u^2(\gamma'(s))^2}{4(1-u\gamma(s))^4}\,,
\]
 % -------------- %
where we use the standard shorthands, $f_{x_j}= \frac{\partial f}{\partial x_j}$. Then
 % -------------- %
\begin{eqnarray}\nonumber
\lefteqn{
\iint\limits_{\Sigma_a}\(|f_{x_1}|^2+|f_{x_2}|^2\)\d x_1\d
x_2-\beta\int\limits_{\Gamma}|f(x)|^2\,\d s} \\
\nonumber && =\int\limits_{\R}\int\limits_0^a
\frac{1}{(1-u\gamma(s))^2}\left|\frac{\partial\varphi}{\partial
s}\right|^2\d u\,\d s+ \int\limits_{\R}\int\limits_0^a
\left|\frac{\partial\varphi}{\partial u}\right|^2\d u\,\d s
\\\nonumber &&+
\int\limits_{\R}\int\limits_0^a V(s,u) \left|\varphi\right|^2\d
u\,\d s
-\int\limits_{\R}\(\frac{\gamma(s)}{2}+\beta\)|\varphi(s,0)|^2\,\d
s\\&&
+\int\limits_{\R}\frac{\gamma(s)}{2(1-a\gamma(s))}\,|\varphi(s,a)|^2\,\d
s\,,\label{quadratic form for q beta}
\end{eqnarray}
 % -------------- %
where
 % -------------- %
\begin{eqnarray}
\lefteqn{V(s,u)=\widetilde
V(s,u)-\frac{\partial}{\partial
s}\(\frac{u\gamma'(s)}{2(1-u\gamma(s))^3}\)-\frac{\partial}{\partial
u}\(\frac{\gamma(s)}{2(1-u\gamma(s))}\)} \nonumber \\ &&
=-\dsfrac{\gamma^2(s)}{4(1-u\gamma(s))^2}-\frac{u\gamma''(s)}{2(1-u\gamma(s))^3}
-\frac{5}{4}\frac{u^2(\gamma'(s))^2}{(1-u\gamma(s))^4}\,. \phantom{AAAAA} \label{effpotential}
\end{eqnarray}
 % -------------- %
Armed with these formul{\ae} we can now introduce the sought pair of estimating operators on $L^2(\mathbb{R}\times(0,a)$. We consider the domains
 % -------------- %
\begin{equation} \label{Qformdomains}
Q_a^{D}=\left\{\varphi\in H^1(\R\times(0,a)):\:
\varphi(.,a)=0 \right\}\,, \quad
Q_a^{N}= H^1(\R\times(0,a))\,,
\end{equation}
 % -------------- %
and define on them the quadratic forms $b_{a,\beta}^{D}[\varphi]$ and $b_{a,\beta}^{N}[\varphi]$, respectively, being equal to the right-hand side of (\ref{quadratic form for q beta}); for $b_{a,\beta}^{D}[\varphi]$, the last summand is skipped.

Using these definitions, it is straightforward to check easily the following claim analogous to Proposition~2.2 of Ref.~\cite{Exner_Yoshitomi 2002}.

 % -------------- %
\begin{lemma} \label{l: uniteq}
The operators $B_{a,\beta}^{D}$, $B_{a,\beta}^{N}$ associated with the above quadratic forms are unitarily equivalent to $L_{a,\beta}^{D}$, $L_{a,\beta}^{N},$ respectively.
\end{lemma}
 % -------------- %

\noindent Our aim is to get bounds to $B_{a,\beta}^{D}$, $B_{a,\beta}^{N}$ just introduced using operators with separated variables. We put
 % -------------- %
\begin{eqnarray*}
&& V_{+}(s):=-\frac{\gamma^2(s)}{4(1+a\gamma_+)^2}+\frac{a\gamma''_+}{2(1-a\gamma_+)^3}\,, \\
&& V_{-}(s):=-\frac{\gamma^2(s)}{4(1-a\gamma_+)^2}-\frac{a\gamma''_+}{2(1-a\gamma_+)^3}
-\frac{5}{4}\frac{a^2(\gamma'_+)^2}{(1-a\gamma_+)^4}\,,
\end{eqnarray*}
 % -------------- %
and estimate the right-hand side of the expression (\ref{quadratic form for q beta}). For an $a$ satisfying $0<a<\gamma_+/2$ and $\varphi$ belonging to $Q_a^D$ and $Q_a^N$, respectively, we define
 % -------------- %
\begin{eqnarray*} \lefteqn{\widetilde
b_{a,\beta}^{D}[\varphi] = (1-a\gamma_+)^{-2}\iint\limits_{0\
\R}^{\ \ \ a\ } \left|\frac{\partial\varphi}{\partial
s}\right|^2\d s\,\d u +
 \iint\limits_{0\ \R}^{\ \ \ a\ }
\left|\frac{\partial\varphi}{\partial u}\right|^2\d s\,\d
u}\\&& +\iint\limits_{0\ \R}^{\ \ \ a\ }V_{+}(s)
\left|\varphi\right|^2\d s\,\d
u-\(\frac{\gamma_*}{2}+\beta\)\int\limits_{\R}|\varphi(s,0)|^2\d
s
\end{eqnarray*}
 % -------------- %
and
 % -------------- %
\begin{eqnarray*}
\lefteqn{\widetilde b_{a,\beta}^{N}[\varphi] =
(1+a\gamma_+)^{-2}\iint\limits_{0\ \R}^{\ \ \ a\ }
\left|\frac{\partial\varphi}{\partial s}\right|^2\d s\,\d u +
\iint\limits_{0\ \R}^{\ \ \ a\ }
\left|\frac{\partial\varphi}{\partial u}\right|^2\d s\,\d u +
\iint\limits_{0\ \R}^{\ \ \ a\ }V_{-}(s) \left|\varphi\right|^2\d
s\,\d
u}\\&& -\(\frac{\gamma^*}{2}+\beta\)\int\limits_{\R}|\varphi(s,0)|^2\,\d
s+\dsfrac{\gamma_*}{2(1-a\gamma_*)}\int\limits_{\R}\left|\varphi(s,a)\right|^2\,\d
s\,; \phantom{AAAAAAAAAAAAAAAAAAAAAAAAAA}
\end{eqnarray*}
% ---------------- %
then we have
 % -------------- %
\begin{equation}\label{inequality b^D}
b_{a,\beta}^{D}[\varphi]\leq \widetilde
b_{a,\beta}^{D}[\varphi]\quad \textrm{ for }\,f\in Q_a^{D},
\end{equation}
 % -------------- %
\begin{equation}\label{inequality b^N}
b_{a,\beta}^{N}[\varphi]\geq \widetilde
b_{a,\beta}^{N}[\varphi]\quad \textrm{ for }\,f\in Q_a^{N}.
\end{equation}
 % -------------- %
Let $\widetilde H_{a,\beta}^{D}$, $\widetilde H_{a,\beta}^{N}$ be the self-adjoint operators associated with the forms $\widetilde b_{a,\beta}^{D}$, $\widetilde b_{a,\beta}^{N}$, respectively. By $T_{a,\beta}^{D}$ we denote the self-adjoint operator associated with the form
 % -------------- %
\[
t_{a,\beta}^{D}[\varphi]=\int\limits_0^a|\varphi'(u)|^2\d
u-\(\frac{\gamma_*}{2}+\beta\)|\varphi(0)|^2
\]
 % -------------- %
defined on $\{\varphi\in H^1(0,a):\: \varphi(a)=0\}$. Similarly, $T_{a,\beta}^{N}$ will be the self-adjoint operator associated with the form
 % -------------- %
\[
t_{a,\beta}^{N}(\varphi,\varphi)=\int\limits_0^a|\varphi'(u)|^2\d
u-\(\frac{\gamma^*}{2}+\beta\)|\varphi(0)|^2+\dsfrac{\gamma_*}{2(1-a\gamma_*)}|\varphi(a)|^2,
\;\; \varphi\in H^1(0,a)\,.
\]
 % -------------- %
Furthermore, we define
 % -------------- %
\[
U_a^{D/N}=-(1\mp a\gamma_+)^{-2}\frac{\d^2}{\d s^2}+V_{\pm}(s) \quad \textrm{ in } L^2(\R)\;\; \textrm{ with the domain }H^2(\R)\,,
\]
 % -------------- %
where the upper/lower sign refers to the index $D/N$, respectively. This allows us to write the estimating operators in the form
 % -------------- %
\begin{equation}\label{H tilde as sum}
\widetilde H_{a,\beta}^{D}=U_a^{D}\otimes I+I\otimes
T_{a,\beta}^{D}\,, \quad \widetilde H_{a,\beta}^{N}=U_a^{N}\otimes
I+I\otimes T_{a,\beta}^{N}\,,
\end{equation}
 % -------------- %
allowing us to assess contributions coming from the longitudinal and transverse variables separately.

Consider first the longitudinal part. We denote by $\mu_j^D(a)$ and $\mu_j^N(a)$ the $j$-th eigenvalue of $U_a^D$ and $U_a^N$, respectively, and recall the result of Proposition~2.3 in Ref.~\cite{Exner_Yoshitomi 2002} (valid for operators on the line corresponding to an infinite boundary curve as well \cite{Exner_Yoshitomi 2001}):

 % -------------- %
\begin{lemma}\label{longlemma}
There exists a constant $C>0$ such that the estimates
 % -------------- %
\begin{equation}\label{estimate mu_j^D}
|\mu_j^D(a)-\mu_j|\leq C a
\end{equation}
 % -------------- %
and
 % -------------- %
\begin{equation}\label{estimate mu_j^N}
|\mu_j^N(a)-\mu_j|\leq C a
\end{equation}
 % -------------- %
hold for any $0<a<(2\gamma_+)^{-1}$, where $C$ is independent of $a,\,j$.
\end{lemma}
 % -------------- %

\noindent On the other hand, Lemmata 2.4 and 2.5 of Ref.~\cite{EMP} allow us to estimate the principal eigenvalue of $T_{a,\beta}^{D}$ and $T_{a,\beta}^{N}$ with an exponential precision. Specifically, we have the following claims:

 % -------------- %
\begin{lemma}\label{lemma T D eigenvalue}
Assume that $\,a\!\left(\beta+\frac{\gamma_*}{2}\right)>\frac{4}{3}.$ Then $T_{a,\beta}^{D}$ has only one negative eigenvalue which we denote
by $\zeta_{a,\beta}^{D}.$ It satisfies the inequalities
 % -------------- %
\[
-\(\beta+\frac{\gamma_*}{2}\)^2\leq\zeta_{a,\beta}^{D}\leq
-\(\beta+\frac{\gamma_*}{2}\)^2+4\(\beta+\frac{\gamma_*}{2}\)^2
\e^{-a\(\beta+\frac{\gamma_*}{2}\)}.
\]
 % -------------- %
\end{lemma}
 % -------------- %
\begin{lemma}\label{lemma T N eigenvalue}
Assume that
$\(\beta+\frac{\gamma^*}{2}\)>\max\left\{\left|\frac{\gamma_*}{2(1-a\gamma_*)}\right|,\
\frac{2\log 5}{3a}\right\}.$ Then $T_{a,\beta}^{N}$ has a unique
negative eigenvalue $\zeta_{a,\beta}^{N},$ and moreover, we have
 % -------------- %
\[
-\(\beta+\frac{\gamma^*}{2}\)^2-\frac{45}{4}\(\beta+\frac{\gamma^*}{2}\)^2
\e^{-a\(\beta+\frac{\gamma^*}{2}\)}\leq\zeta_{a,\beta}^N\leq
-\(\beta+\frac{\gamma^*}{2}\)^2.
\]
 % -------------- %
\end{lemma}
 % -------------- %

\noindent Now we are in position to prove the inequalities \eqref{estimate_lambda n} for the first $M$ eigenvalues \emph{provided they exist}. The bracketing estimate we have squeezes the eigenvalues in question between those of the operators \eqref{H tilde as sum}. Since the latter have separated variables, their eigenvalues are sums of eigenvalues of the longitudinal and transverse component which can be estimated using Lemmata~\ref{longlemma} and \ref{lemma T N eigenvalue}, hence it is sufficient to choose $a= \frac{3}{\beta} \log\beta$ to arrive at the inequalities \eqref{estimate_lambda n}.

At the same time, these estimates does not help to establish the
existence of the eigenvalues, because the essential spectrum
thresholds of the operators \eqref{H tilde as sum} are wide apart.
Indeed, the essential spectrum of $U_a^{D/N}$ starts in view of
the assumption (iii) at zero, hence Lemmata~\ref{longlemma},
~\ref{lemma T D eigenvalue}, \ref{lemma T N eigenvalue} yield
 % -------------- %
\begin{eqnarray*} && \inf\sigma_\mathrm{dis}(\widetilde
H_{a,\beta}^{D}) = -\beta^2-\gamma_* \beta + \OO\(1\)\,,
\\
&& \inf\sigma_\mathrm{ess}(\widetilde H_{a,\beta}^{N}) =
-\beta^2-\gamma^*\beta + \OO\(1\)\,.
\end{eqnarray*}
 % -------------- %
Consequently, we cannot be sure that the eigen\-values of
$\widetilde H_{a,\beta}^{D}$ are situated below the essential
spectrum threshold of the original operator. To overcome this
difficulty we need to derive a better upper estimate of the
operator $B_{a,\beta}^D$ from Lemma~\ref{l: uniteq}. First we note
that its essential spectrum threshold is close to $-\beta^2$.

 % -------------- %
\begin{lemma}\label{lemma B_D infess}
Under our assumptions we have
 % -------------- %
\begin{equation}\label{B_D infess}
\inf\sigma_\mathrm{ess}(B_{a,\beta}^{D}) = -\beta^2 + \OO\(\frac{\log\beta}{\beta}\)\,.
\end{equation}
 % -------------- %
\end{lemma}
 % -------------- %
\emph{Proof:} We employ once more Dirichlet-Neumann bracketing and squeeze $B_{a,\beta}^{D}$ between a pair of operators on $L^2(\mathbb{R}\times(0,a)$, % -------------- %
\begin{equation}\label{infess_est}
B_{a,\beta,s_0}^{DN} \le B_{a,\beta}^{D} \le B_{a,\beta,s_0}^{DD}\,,
\end{equation}
 % -------------- %
obtained from $B_{a,\beta}^{D}$ by adding Neumann and Dirichlet condition, respectively, at the segments $\pm s_0\times(0,a)$. Each of the estimating operators is a direct sum of three parts. The middle one refers to a precompact region, hence it does not contribute to the essential spectrum, hence it is sufficient to assess the tail parts. This can be done in a way analogous to \eqref{H tilde as sum}. The difference is that now we consider the curvature only for $|s|>s_0$, thus in the transverse part we modify Lemmata~\ref{lemma T D eigenvalue}, \ref{lemma T N eigenvalue} by replacing $\gamma^*$ and $\gamma_*$ by the maximum and minimum of $\gamma(s)$ in the tail regions; in view of assumption (iii) the moduli of these quantities can be made arbitrarily small by choosing $s_0$ large enough. The same applies to the contribution of the first terms in $V_\pm(s)$ to the longitudinal part, while the remaining ones are in view of assumption (iv) proportional to $a$ giving rise to the error term in \eqref{infess_est}. \hfill $\Box$

 % -------------- %
\begin{remark} \label{rem: better squeeze}
{\rm If we strengthen assumption (iv) requiring, in addition, that $\lim_{|s|\to\infty} \gamma^{(j)}=0$ for $j=1,2$, we can localize the essential spectrum threshold with an exponential precision, however, we do not need such a claim to prove Theorem \ref{main theorem_half-plane}.}
\end{remark}
 % -------------- %

\noindent Now we can complete the proof by replacing the upper bound coming from (\ref{H tilde as sum}) by a more precise variational estimate of the operator $B_{a,\beta}^D$. Consider first its principal eigenvalue which satisfies $\lambda_1(\beta)\|\varphi\|^2_{L^2}\le b_{a,\beta}^D[\varphi]$ for any $\varphi\in Q_a^D$ and construct the following family of trial functions,
 % -------------- %
$$
\hat\varphi(s,u)=\chi_{\varepsilon}(s)\(\e^{-\alpha u}-\e^{-2a\alpha+u\alpha}\)\,,
$$
 % -------------- %
where $\chi_{\varepsilon}$ is a smooth function on $\R$ with the support located in an $\varepsilon$-neighborhood of a point $s^*$ in which the curvature reaches its maximum, $\gamma(s^*)=\gamma^*$, and $\varepsilon$ is a parameter to be determined later. In view of the smoothness of $\gamma$ in combination with assumption (iii), at least one such point exists. The function $\chi_{\varepsilon}$ used above is supposed to be of the form
 % -------------- %
\[
\chi_{\varepsilon}(s):=\chi\(\dsfrac{s-s^*+\varepsilon}{2\varepsilon}\)\,,
\]
 % -------------- %
where $\chi(x)$ is a fixed smooth function on $\mathbb{R}$ with the support in the interval $(0,1)$. It is straightforward to check the scaling relations,
 % -------------- %
\begin{equation}\label{norm chi}
\|\chi_{\varepsilon}\|_{L^2(\R)}^2=2\varepsilon\|\chi\|_{L^2(0,1)}^2,\quad
\|\chi'_{\varepsilon}\|_{L^2(\R)}^2=\(2\varepsilon\)^{-1}\|\chi'\|_{L^2(0,1)}^2.
\end{equation}
 % -------------- %
We also note that on the support of $\chi_{\varepsilon}$, i.e. for any $s\in(s^*-\varepsilon,s^*+\varepsilon)$ we have
 % -------------- %
\[|\gamma(s)-\gamma^*|<\gamma'_+ |s-s^*| < \gamma'_+ \varepsilon.\]
 % -------------- %
Computing the terms of the form $b_{a,\beta}^D[\varphi]$ we get for the longitudinal kinetic contribution the estimate
 % -------------- %
\begin{eqnarray*}
\lefteqn{\hspace{-1.5em} \iint\limits_{0\ \R}^{\ \ \ a\ }
\frac{1}{(1-u\gamma(s))^2}\left|\frac{\partial\hat\varphi}{\partial
s}\right|^2\d s\,\d u \le \int\limits_{0}^a\int\limits_{
s^*-\varepsilon}^{s^*+\varepsilon}
\(\frac{1}{(1-u\gamma^*)^2}+c\varepsilon u\)\(\e^{-\alpha
u}-\e^{-2a\alpha+u\alpha}\)^2} \\ &&  \times (\chi'(s))^2\d s\,\d u
\left[\(\dsfrac{1}{2\alpha}+\OO\big(\alpha^{-2}\)\big)+c\varepsilon\(\dsfrac{1}{4\alpha^2}+
\OO\big(\alpha^{-3}\big)\)\right]\|\chi\|^2_{L^2(\R)}\,,
\end{eqnarray*}
 % -------------- %
where $c>0$ is a generic constant independent of $\beta, a$, and $\varepsilon.$ Similarly,
 % -------------- %
\[
\iint\limits_{0\ \R}^{\ \ \ a\ }
\left|\frac{\partial\hat\varphi}{\partial u}\right|^2\d s\,\d
u=\dsfrac{\alpha}{2}\left(1+\OO\big(\alpha\e^{-2a\alpha}\big)\right)\|\chi\|_{L^2(0,L)}^2
\]
 % -------------- %
holds for the transverse kinetic term,
 % -------------- %
\begin{eqnarray*}
\lefteqn{\int\limits_{0}^a\int\limits_{ \R}^{}V(s,u)
\left|\hat\varphi\right|^2\d s\,\d u} \\ && \le
\int\limits_{0}^a\int\limits_{
s^*-\varepsilon}^{s^*+\varepsilon}\(-\dsfrac{\(\gamma^*\)^2}{4(1-u\gamma^*)}
-\dsfrac{u\gamma''(s^*)}{2(1-u\gamma^*)^3}-\dsfrac{5}{4}\dsfrac{u^2\(\gamma'(s^*)\)^2}
{\(1-u\gamma^*\)^4}+c\varepsilon\) \\ && \times
\(\e^{-\alpha u}-\e^{-2a\alpha+u\alpha}\)^2 |\chi(s)|^2\d s\d u \\ &&
=\(-\dsfrac{\(\gamma^*\)^2}{4}+c\varepsilon\)
\dsfrac{1}{2\alpha}\(1+\OO\big(\alpha^{-1}\big)\)\|\chi\|^2_{L^2(\R)}
\end{eqnarray*}
 % -------------- %
for the potential one, and
 % -------------- %
\[
-\int\limits_{\R}^{}\(\frac{\gamma(s)}{2}+\beta\)|\hat\varphi(s,0)|^2\d s\geq-\(\beta+\dsfrac{\gamma^*-\varepsilon}{2}\)
\(1-\e^{-2a\alpha}\)^2\|\chi\|_{L^2(\R)}^2,
\]
 % -------------- %
for the boundary one. Finally, the trial function norm satisfies
 % -------------- %
\[\iint\limits_{0\ \R}^{\ \ \ a\ }|\hat\varphi(s,u)|^2\d s\d u=\dsfrac{1}{2\alpha}\(1+\OO\big(\alpha\e^{-2a\alpha}\big)\)\|\chi\|^2_{L^2(\R)}.\]
 % -------------- %
Putting these expressions together and taking (\ref{norm chi}) into account we get
 % -------------- %
\begin{eqnarray*}
\lefteqn{\dsfrac{b_{a,\beta}^D[\hat\varphi]}{\|\hat\varphi\|^2_{L^2(\R)}}\le
\dsfrac{1}{4\varepsilon^2}\
\dsfrac{\|\chi'\|_{L^2(0,1)}^2}{\|\chi\|_{L^2(0,1)}^2}\(1+\OO\big(\alpha^{-1}\big)+c\varepsilon\(\dsfrac{1}{2\alpha}+\mathrm{\alpha^{-2}}\)\)}
\\ && +\alpha^2\(1+\OO\big(\alpha\e^{-2a\alpha}\big)\) +
\(-\dsfrac{\(\gamma^*\)^2}{4}+c\varepsilon\)\(1+\OO\big(\alpha^{-1}\big)\)
\\ &&
-2\alpha\(\beta+\dsfrac{\gamma^*-\varepsilon}{2}\)\(1+\OO\big(\alpha\e^{-2a\alpha}\big)\).
%\label{ineq 1}
\end{eqnarray*}
 % -------------- %
Now we choose $\alpha=\beta+\dsfrac{\gamma^*}{2}$ in which case the right-hand side of the last inequality becomes
 % -------------- %
\begin{eqnarray*}
\lefteqn{\hspace{-2em} \dsfrac{1}{4\varepsilon^2}\
\dsfrac{\|\chi'\|_{L^2(0,1)}^2}{\|\chi\|_{L^2(0,1)}^2}\(1+\OO\big(\beta^{-1}\big)+c\varepsilon\(\dsfrac{1}{2\beta}+\mathrm{\beta^{-2}}\)\)
-\(\beta+\dsfrac{\gamma^*}{2}\)^2} \\ && +\varepsilon\(\beta+\dsfrac{\gamma^*}{2}\)
+\(-\dsfrac{\(\gamma^*\)^2}{4}+c\varepsilon\)\(1+\OO\big(\alpha^{-1}\big)\),
\end{eqnarray*}
 % -------------- %
and to optimize the last formula with respect to $\varepsilon$ we take $\varepsilon=\beta^{-1/3},$ which yields the estimate
 % -------------- %
\begin{equation} \label{finalbound}
\dsfrac{b_{a,\beta}^D[\hat\varphi]}{\|\hat\varphi\|^2_{L^2(0,L)}}\le-\(\beta+\dsfrac{\gamma^*}{2}\)^2+\OO\big(\beta^{2/3}\big).
\end{equation}
 % -------------- %
Since by Lemma~\ref{lemma B_D infess} the window given by this estimate does not overlap with the essential spectrum, the operator $B_{a,\beta}^D$ has an isolated eigenvalue, and the same is by Lemma~\ref{l: uniteq} and the min-max principle, Ref.~\cite[Sec.~XIII.1]{Reed Simon 4 1978}, true for the operator $L_{a,\beta}^D$, and in turn also for the original operator $H_\beta$.

The argument for the higher eigenfunctions proceeds in the same way. In this case we employ slightly modified trial functions of the form
 % -------------- %
$$
\hat\varphi_j(s,u)=\chi_{\varepsilon,j}(s)\(\e^{-\alpha u}-\e^{-2a\alpha+u\alpha}\)\,,
$$
 % -------------- %
where the longitudinal part contains the function $\chi_{\varepsilon}$ with a shifted argument, for instance
 % -------------- %
\[
\chi_{\varepsilon,j}(s):=\chi\(\dsfrac{s-s^*+(2j-1)\varepsilon}{2\varepsilon}\)\,.
\]
 % -------------- %
The above estimate of the form remains essentially the same, up to the values of the constants involved. By construction, the functions $\chi_{\varepsilon,j}$ with different values of~$j$ have disjoint supports, hence $\hat\varphi_j$ is orthogonal to $\hat\varphi_i,\: i=1,\dots,j-1,$ and using once more the min-max principle we conclude that the eigenvalue $\lambda_j(\beta)$ is again isolated for $\beta$ large enough and has the upper bound given by the right-hand side of (\ref{finalbound}). Choosing $j=N$, we obtain in combination with \eqref{estimate_lambda n} the first claim of the theorem, the inequality \eqref{estimate_lambda n_} follows from the fact that the eigenvalues are arranged in the ascending order. \hfill $\Box$

%%%%%%%%%%%%%%%%%%%%%%%%%%%%%%%%%%%%%%%%%%%%%%%%%%%%%%%
\section{Bound state asymptotics of a curved Robin waveguide}\label{sect: waveguide}

Since the analysis performed so far was based on behavior of the solutions in the vicinity of the boundary, most of the above results can be extended to situations when the boundary has several disjoint components. A case of particular interest concerns the case when $\Omega$ is a (non-straight) strip of a constant width $d>0$. While in the case when the boundary is a single curve the discrete spectrum may be empty --- for instance, if $\Omega$ is strictly concave --- a Robin waveguide has always isolated eigenvalues unless it is straight, similarly to the Dirichlet one treated in Ref.~\cite{Exner_Seba 1989}. The claim was proved in Ref.~\cite{Jilek_06} for repulsive Robin boundary, $\beta<0$, but the argument carries over without any modification to the attractive case; the only exception is the Neumann case, $\beta=0$. In fact, the proof in Ref.~\cite{Jilek_06} is done under the assumption that the strip in straight outside a compact region, however, it is easily modified to cover situation when it is straight only asymptotically.

Consider one boundary of the strip as reference one, called $\Gamma_0$ characterized by its curvature $\gamma_0$, and suppose that that it satisfies assumptions (ii)--(iv) of Sec.~\ref{sect: half-plane}. The strip we are interested in can be regarded as the family of `parallel' curves
$\Gamma_u:\: \R\ni s\mapsto (\Gamma_{0,1}(s)-u\Gamma_{0,2}'(s),\Gamma_{0,2}(s)+u\Gamma_{0,1}'(s))\in\mathbb{R}^2$ with $u\in (0,d)$, while $\Gamma_d$ is its other boundary. The curvature of $\Gamma_u$ can be expressed as $\gamma_u(s)=\frac{\gamma(s)}{1-u\gamma(s)}$. We introduce the quantities $\gamma_0^*$ and $\gamma_{0,*}$ analogous to the extreme of $\gamma$ in \eqref{curv_extreme} and
 % -------------- %
$$
\gamma_d^*:=\max\limits_{s\in\R}\gamma_d(s)=\frac{\gamma^*}{1-d\gamma^*}\,,\quad \gamma_{d,*}:=\min\limits_{s\in\R}\gamma_d(s)=\frac{\gamma_*}{1-d\gamma_*}\,.
$$
 % -------------- %
Since the width $d$ introduces a length scale into the problem we replace the assumption (i) by the following requirement:

 % -------------- %
\begin{description}
\item (i') $\:\max\{d\gamma_0^*,\, d\gamma_d^*\}<1$ and the strip $\Omega$ does not intersect itself.
\end{description}
 % -------------- %

\noindent The first requirement guarantees the existence of the `straightening' transformation analogous to that of Lemma~\ref{lemma_existence of C_u} locally, the second one globally. As before we are interested in the boundary-value problem
 % -------------- %
\begin{eqnarray}\label{Robin_problem waveguide}
-\Delta f=\lambda f\; &\textrm {in}& \Omega \nonumber \\ [-.5em]
&& \label{problem} \\ [-.5em] \frac{\partial f}{\partial n}=\beta
f\; &\textrm{on}& \Gamma:=\Gamma_0\cup \Gamma_d \nonumber
\end{eqnarray}
 % -------------- %
with a parameter $\beta>0$ assumed to be large. The normal has at both part of the boundary the outward direction; note that the parametrization of $\Gamma_d$ by its arc length has the opposite orientation.

The argument of the previous section can be copied \emph{verbatim} for each boundary component and $a<\frac12 d$. The spectrum is then estimated by the union of the spectra coming from the strip neighborhoods of $\Gamma_0$ and $\Gamma_d$. Due to the opposite orientation, the curvature to consider for the latter is in fact $-\gamma(s)$. As at least one of the quantities $\gamma_0^*,\, -\gamma_{d,*}$ is positive, unless $\Omega$ is straight, we get the existence of arbitrarily many bound states in asymptotic regime, $\beta\rightarrow\infty$. In particular, Theorem~\ref{main theorem_half-plane} implies the following claim:

% --------- %
\begin{theorem} \label{waveguide theorem}
Suppose that $\Omega$ is not straight and adopt the assumption (i'), (ii)--(iv). Then to any positive $N$ there is a $\beta_N$ such that for any
$\beta>\beta_N$ the Robin Laplacian $H_\beta$ on $\Omega$ has at least $N$ eigenvalues with the following asymptotic expansion,
 % -------------- %
$$
\lambda_{j}(\beta) = -\beta^2-\max\left\{\gamma^*,-\gamma_{d,*}\right\}\beta+\OO\big(\beta^{2/3}\big)\,,\quad j=1,\dots,N\,.
$$
 % -------------- %
\end{theorem}

%%%%%%%%%%%%%%%%%%%%%%%%%%%%%%%%%%%%%%%%%%%%%%%%%%%%%%%
\section{Bound states in the non-asymptotic regime}
\label{sect: Proof of existence of bound state}

Let us return now to our basic example in which the boundary of $\Omega$ is a single infinite curve. As indicated in the introduction, the results obtained in Sec.~\ref{sect: Proof half plane} motivate us to ask about the existence of bound states beyond the asymptotic regime. While the general setting is the same as before, the spectral properties have now a more global character which forces us to modify the assumptions made in Sec.~\ref{sect: half-plane}. For the sake of simplicity we are going to suppose here that the boundary $\Gamma:\: \R\to\R^2$ of $\Omega$, assumed again to be a $C^4$-smooth curve, is straight outside a compact which opens the following three possibilities:

\begin{enumerate}[(i)]
\setlength{\itemsep}{-3pt}
 % -------------- %
\item $\Gamma$ is a nontrivial local deformation of the straight line, i.e. there is a positive $s_0$ such that $\Gamma_1(s) = \Gamma_1(\pm s_0)+(s\mp s_0)$ and $\Gamma_2(s)=0$ holds for any $\pm s\geq s_0$,
 % -------------- %
\item $\Omega$ is a nontrivial local deformation of a wedge, i.e. there is an $\alpha\in (0,\frac12\pi)$ and a positive $s_0$ such that
 % -------------- %
\begin{equation} \label{wedge}
\left\{\begin{array}{l}
\Gamma_1(s) = \Gamma_1(\pm s_0) + (s\mp s_0)\cos\alpha
\\
\Gamma_2(s) = \Gamma_2(\pm s_0) \pm (s\mp s_0)\sin\alpha
\end{array}\right.
\end{equation}
 % -------------- %
holds for any $\pm s\geq s_0$,
 % -------------- %
\item there is an $\alpha\in [-\frac12\pi,0)$ and an $s_0>0$ such that relations \eqref{wedge} hold for any $\pm s\geq s_0$.
 % -------------- %
\end{enumerate}
 % -------------- %

\noindent Note that the assumptions (i) and (iii) partly overlap, the latter with $\alpha=0$ covers also the situation when the `outer' components of the boundary are parallel but not necessarily parts of a single line.

First we shall identify the essential spectrum of the corresponding Robin Laplacian $H_\beta$ associated with the boundary-value problem (\ref{problem_0}) in a way which strengthens under the present assumptions the claim of Lemma~\ref{lemma B_D infess}.

 % -------------- %
\begin{theorem}\label{thm: continuous spectrum}
Any of the assumptions (i)--(iii) implies $\sigma_\mathrm{ess}(H_\beta)=[-\beta^2,\infty)$.
\end{theorem}
% -------------- %
\emph{Proof:}
Let us check first that $[-\beta^2,\infty)\subset\sigma_\mathrm{ess}(H_{\beta})$. To this end it is sufficient to construct an appropriate Weyl sequence, i.e. to find $f_n\in D(H_{\beta}) = \left\{f\in H^2(\Omega):\ \frac{\partial f}{\partial n} = \beta f\right\}$ such that $\|f_n\|_{L^2(\Omega)} = 1$ and
 % -------------- %
$$
\|H_{\beta,0}f_n-\lambda f_n\|_{L^2(\Omega)} =
\|-(f_n)_{xx}-(f_n)_{yy}-\lambda f_n\|_{L^2(\Omega)}
\rightarrow 0
$$
 % -------------- %
holds as $n\to\infty$ for any fixed $\lambda\in[-\beta^2,\infty)$. Spectral properties of $H_\beta$ are certainly invariant w.r.t. the choice of the Cartesian system in the plane. We thus rotate the domain $\Omega$ by the angle $\alpha$ clockwise; then $\Gamma$ will contain the segment $[x_0,\infty)$ of the real axis for some number $x_0$. We choose a function $\varphi\in C_0^{\infty}$ such that $\mathrm{supp}\,\varphi\subset(-1,1)$ and $\|\varphi\|_{L^2(\R)} = 1$ and define $\varphi_n(x) := \frac{1}{\sqrt{n}}\varphi\(\frac{x}{n}-n\)$. Then $\mathrm{supp}\,\varphi_n \subset(n^2-n,n^2+n)$, the function is normalized, $\|\varphi_n\|_{L^2(\R)} = 1$, and the norms $\|\varphi'_n\|_{L^2(\R)}$ and $\|\varphi''_n\|_{L^2(\R)}$ vanish as $n\rightarrow\infty$.

Let us now put $f_n(x,y) = \sqrt{2\beta}\,\varphi_n(x)\,\e^{\i\sqrt{\lambda+\beta^2}\, x}\ \e^{-\beta y}$. These functions are for $\alpha\in [-\frac12\pi,\frac14\pi]$ and all $n$ large enough normalized as needed,
 % -------------- %
\[
\|f_n\|^2_{L^2(\Omega)}=\int\limits_{\Omega} |f_n(x,y)|^2\,\d x \d y =
\int\limits_{n^2-n}^{n^2+n}\int\limits_{0}^{\infty}
2\beta\,|\varphi_n(x)|^2\,\e^{-2\beta y}\,\d x \d y =1\,,
\]
 % -------------- %
they belong to $D(H_{\beta})$ and satisfy
 % -------------- %
\begin{equation} \label{Weylnorm}
\(H_{\beta,0}f_n-\lambda f_n\)(x,y) = -
\sqrt{2\beta}\(\varphi''_n(x)+2\i\sqrt{\lambda+\beta^2}\ \varphi'_n(x)\)
\e^{\i\sqrt{\lambda+\beta^2}\,x}\ \e^{-\beta y},
\end{equation}
 % -------------- %
hence $\|H_{\beta}f_n-\lambda f_n\|_{L^2(\R)}\to 0$ holds as
$n\to\infty$. If $\alpha\in (\frac14\pi,\frac12\pi)$ the argument
is no longer valid because the above trial function does not
satisfy the correct boundary conditions at the other part of the
boundary being far away from the origin a halfline of the angle
$\theta:=\pi-2\alpha\in(0,\pi/2)$ with respect to the positive
$x$-half-axis. In that case we choose a function $g\in
C_0^\infty(\R)$ such that
 % -------------- %
\[
g(y)= \left\{ \begin{array}{lcc} 1 &\;\dots\;& 0\le y\le 1 \\[.5em]
0 &\;\dots\;& y\ge 2 \end{array} \right.
\]
 % -------------- %
and put
 % -------------- %
\[
f_n(x,y) =
\sqrt{2\beta}\,\varphi_n(x)\,\e^{\i\sqrt{\lambda+\beta^2}\, x}\
\e^{-\beta y} g\left( \frac{3y}{n^2\tan \theta} \right)\,;
\]
 % -------------- %
it is obvious that these function satisfy trivially the boundary conditions for all $n$ large enough. Their squared norms are given by
 % -------------- %
\[
\|f_n\|^2_{L^2(\Omega)} = \int_0^{\frac23 \beta n^2\tan \theta}
\e^{-u}\,\d u + \int_{\frac23 \beta n^2\tan \theta}^{\frac43 \beta
n^2\tan \theta} \e^{-u}\,g\left( \frac{3u}{2\beta n^2\tan \theta}
\right)^2 \d u\,,
\]
 % -------------- %
hence $\|f_n\|^2_{L^2(\Omega)} = 1 - \OO\big(\e^{-\frac23 \beta
n^2\tan \theta} \big)$. On the other hand, the relation
\eqref{Weylnorm} remains valid for $0\le y\le \frac13 n^2\tan
\theta$, while for $\frac13 n^2\tan \theta \le y \le \frac23
n^2\tan \theta$ its right hand side is multiplied by
$g\(\frac{3y}{ n^2\tan \theta}\)$ and the additional terms
 % -------------- %
\begin{eqnarray*}
\lefteqn{-\sqrt{2\beta} \ \varphi_n(x) \bigg( -2\beta
\frac{3}{n^2\tan \theta} g'\left( \frac{3y}{n^2\tan \theta}
\right)}
\\ && + \frac{9}{n^4\tan^2 \theta} g''\left( \frac{3y}{n^2\tan \theta} \right) \bigg)
\e^{\i\sqrt{\lambda+\beta^2}\,x}\ \e^{-\beta y}
\end{eqnarray*}
 % -------------- %
have appear, and for $y \ge \frac23 n^2\tan \theta$ the expression vanishes; using these observations it is again easy to construct an appropriate Weyl sequence. This shows that $[-\beta^2,\infty) \subset \sigma_\mathrm{ess}(H_{\beta})$.

\begin{figure}
\begin{minipage}[ht!]{0.49\linewidth}
\begin{center}
\epsfig{width=60mm,figure=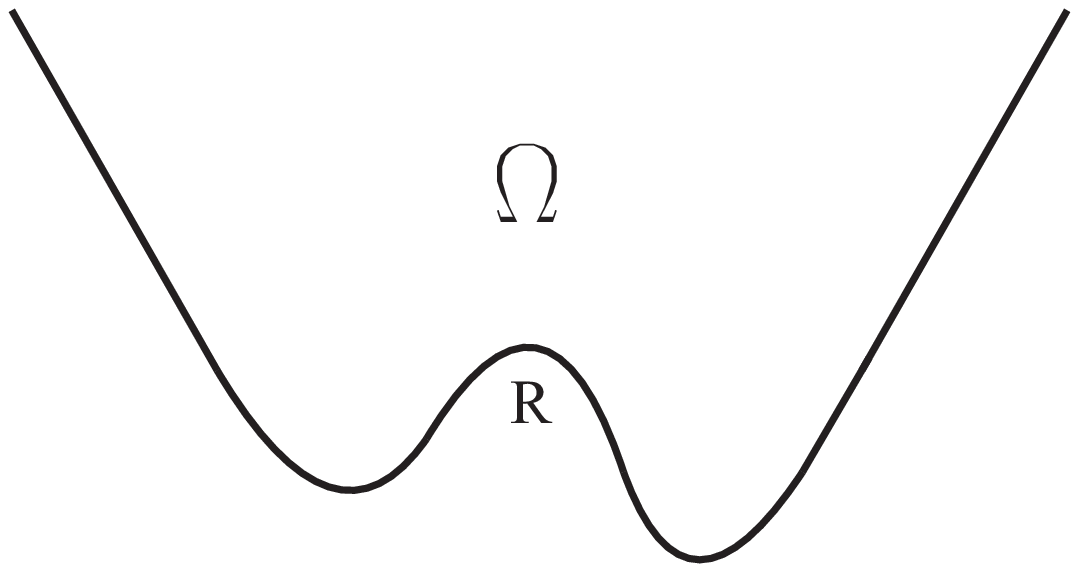}\\Robin boundary
conditions
\end{center}
\end{minipage}
 % ----------------- %
\begin{minipage}[ht!]{0.49\linewidth}
\begin{center}
\epsfig{width=60mm,figure=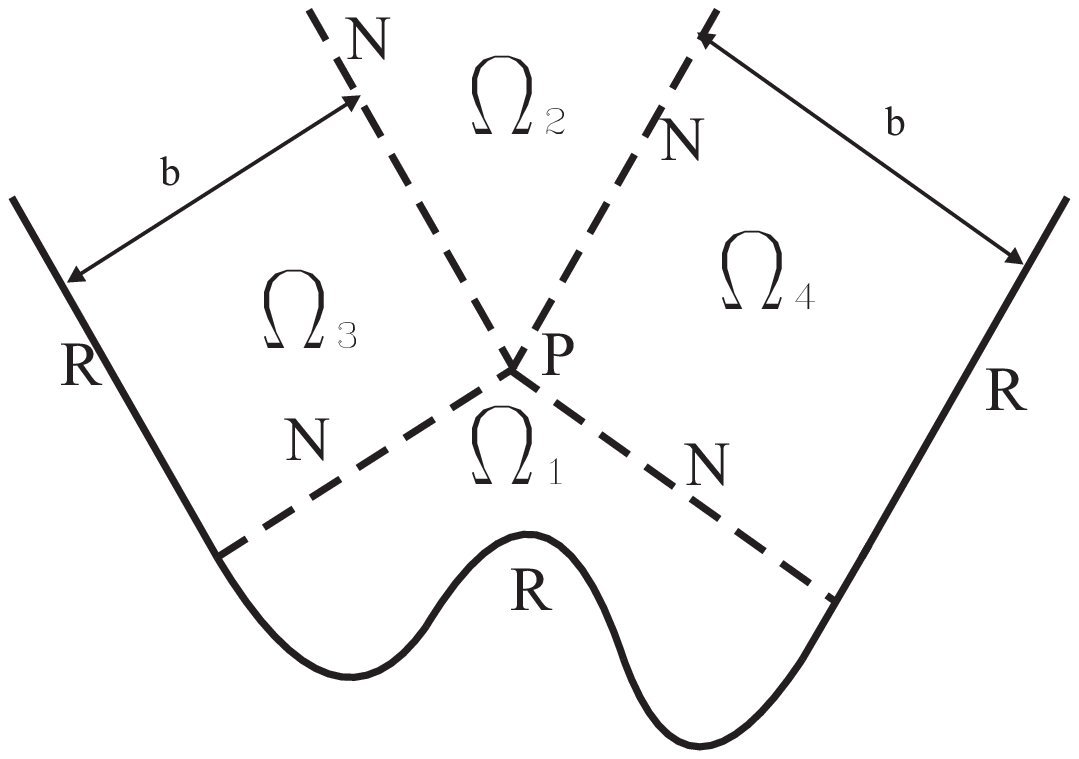}\\Robin-Neumann b.c.
\end{center}
\end{minipage}
\caption{Domain splitting for $\alpha>0$.} \label{Picture: Omega_0}
\end{figure}

\begin{figure}
\begin{minipage}[ht!]{0.49\linewidth}
\begin{center}
\epsfig{width=60mm,figure=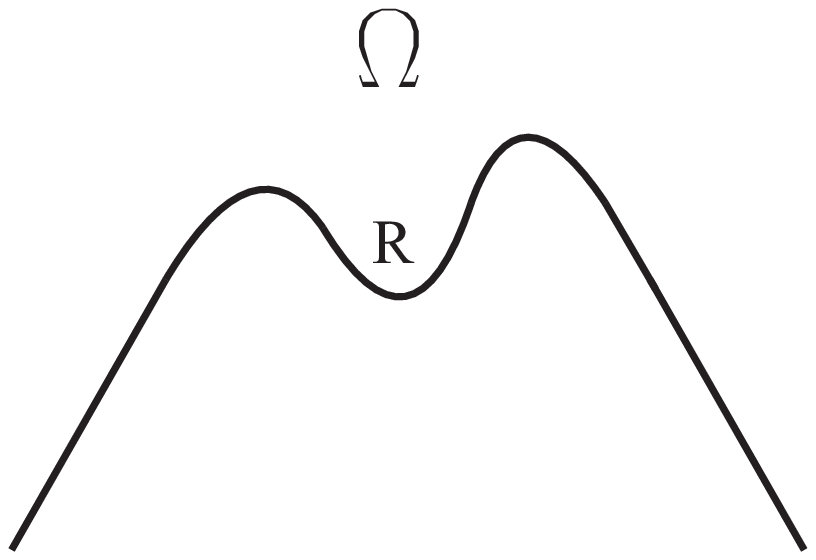}\\ Robin boundary
conditions
\end{center}
\end{minipage}
 % ----------------- %
\begin{minipage}[ht!]{0.49\linewidth}
\begin{center}
\epsfig{width=60mm,figure=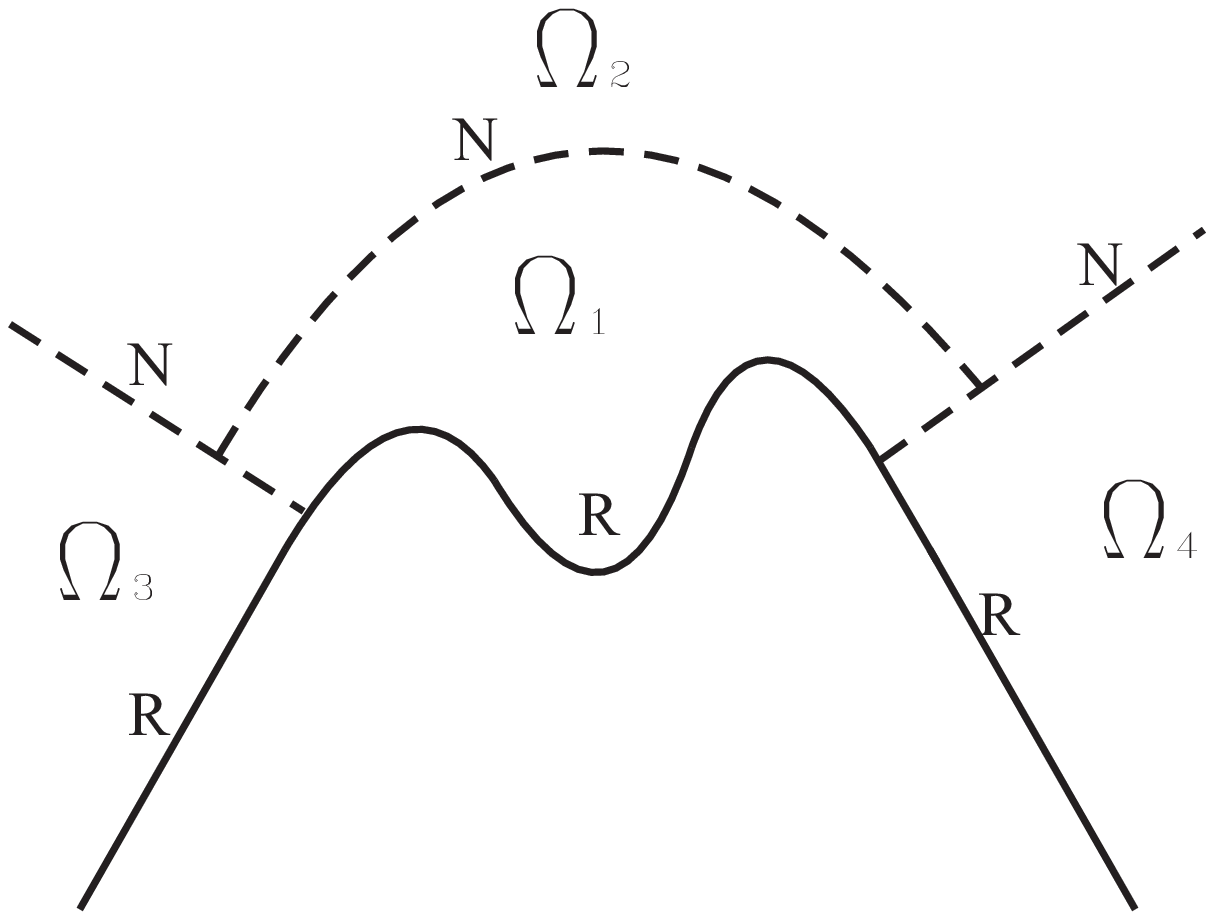}\\ Robin-Neumann b.c.
\end{center}
\end{minipage}
\caption{Domain splitting for $\alpha\le 0$.} \label{Picture: Omega_0 2}
\end{figure}

To prove the opposite inclusion, we use again a bracketing argument dividing the domain $\Omega$ into smaller a union of subdomains $\Omega_j,$ $j=1,2,3,4$, and their boundaries in two different ways as shown in Figures~\ref{Picture: Omega_0} and \ref{Picture: Omega_0 2}, the former referring to $\alpha\in(0,\frac12\pi)$, the latter to $\alpha\in[-\frac12\pi,0]$. Imposing Neumann conditions at the added boundaries, we estimate $H_{\beta}$ from below,
 % -------------- %
\begin{equation}\label{estimate H beta by N}
H_{\beta,0}\geq
\(-\Delta^{NR}_{\Omega_1}\)\oplus\(-\Delta^N_{\Omega_2}\)\oplus
\(-\Delta^{NR}_{\Omega_3}\)\oplus\(-\Delta^{NR}_{\Omega_4}\)\,.
\end{equation}
 % -------------- %

 % -------------- %
 % ------------ %

 % -------------- %
 % -------------- %

\noindent The spectrum of the operator at the right-hand side is, of course, the union the component spectra. The domain $\Omega_1$ is compact and thus it does not contribute to the essential spectrum, the domain $\Omega_2$ does but the corresponding operator is positive by definition.
Hence it is sufficient to find the essential spectrum for the domains $\Omega_3$ and $\Omega_4$ being semi-infinite strips or planar quadrants depending on the sign of $\alpha$; in the former case it is important that the strip width $b$ can be made as large as one wishes.

Denote the corresponding operators by $T^\mathrm{strip}_{\beta,b}$ and $T^\mathrm{quadrant}_{\beta}$; the quadratic forms associated with them
are
 % -------------- %
\[
q^\mathrm{strip}_{\beta,b} = \int\limits_{\R}\int\limits_0^b \(|f_x|^2+|f_y|^2\) \d x\d y -\beta\int\limits_{\R} |f(x,0)|^2\,\d x
\]
 % -------------- %
and
 % -------------- %
\[
q^\mathrm{quadrant}_{\beta,b} = \int\limits_{\R}\int\limits_{\R} \(|f_x|^2+|f_y|^2\) \d x\d y -\beta\int\limits_{\R} |f(x,0)|^2\,\d x\,,
\]
 % -------------- %
respectively. Since the variables decouple the spectra are easily found, in particular, we have $\sigma(T^\mathrm{quadrant}_{\beta}) = \sigma_\mathrm{ess}(T^\mathrm{quadrant}_{\beta}) = [-\beta^2,\infty)$. In the strip case we have to find the spectral threshold of
$-\Delta^\mathrm{NR}_{\beta,(0,b)}$, the Robin-Neumann Laplacian on $(0,b)$ with the domain
 % -------------- %
\[
\left\{f\in H^2(0,b):\ -f'(0)=\beta f(0),\ f'(b)=0\right\}\,.
\]
 % -------------- %
It is straightforward to check that $\inf\sigma(-\Delta^\mathrm{NR}_{\beta,(0,b)}) = -\zeta^2$, where $\zeta = \zeta(b)>0$ is the solution to the equation
 % -------------- %
\[
\frac{\zeta-\beta}{\zeta+\beta} = \e^{-2\zeta b}\,.
\]
 % -------------- %
Since $\lim\limits_{b\rightarrow+\infty}\zeta(b) = \beta$, we conclude that $\inf \sigma_\mathrm{ess}(T^\mathrm{strip}_{\beta}) > -\beta^2 - \varepsilon$ for an arbitrary $\varepsilon>0$, which yields the desired result. \hfill $\Box$

On the other hand, for the existence of a discrete spectrum the sign of the asymptotic bending angle $\alpha$ is important.

 % -------------- %
\begin{theorem}\label{thm: bound states}
The operator $H_{\beta}$ associated with the eigenvalue problem (\ref{problem_0}) has a bound state under any of assumptions (i) or (ii).
\end{theorem}
 % -------------- %
\emph{Proof:}
We start with assumption (i). The quadratic form (\ref{differential form}) associated with the operator $H_{\beta}$ can be written as
 % -------------- %
$$
q_{\beta}[f]= \iint\limits_{\Omega} \(\left| f_x
\right|^2+\left| f_y \right|^2\) \d x\,\d y-\beta
\int\limits_{\Gamma} \left|f\right|^2\d s\,.
$$
 % -------------- %
By the previous theorem the essential spectrum coincides with the interval $[-\beta^2,+\infty)$, hence it is sufficient to find a
function $f\in H^1(\Omega)$ such that
 % -------------- %
$$
S[f]:=q_{\beta}[f]+\beta^2 \|f\|^2_{L^2(\Omega_0)}<0\,.
$$
 % -------------- %
To this aim we choose a function $\psi_1\in C_0^{\infty}(R)$ with the properties similar to that of $g$ of the preceding proof, but two-sided,
 % -------------- %
\[
\psi_1(x)= \left\{ \begin{array}{lcc} 1 &\;\dots\;& 0\le |x|\le 1 \\[.5em]
0 &\;\dots\;& |x|\ge 2 \end{array} \right.
\]
 % -------------- %
and define
 % -------------- %
\begin{equation}\label{f_n}
f_n(x,y)=\psi_n(x)\e^{-\beta y}\,,
\end{equation}
 % -------------- %
where $\psi_n(x):=\psi_1\(\frac{x}{n}\)$; we note that $\|\psi'_n\|_{L^2(\R)}\rightarrow0$ and $\psi_n(x)\rightarrow 1$ pointwise as $n\to\infty$. Then $S[f_n]$ acquires the form
 % -------------- %
$$
\iint\limits_{\Omega} \left|\psi'_n(x)\right|^2\e^{-2\beta
y}\,\d x\,\d y+ 2\beta^2\iint\limits_{\Omega}
\left|\psi_n(x)\right|^2\e^{-2\beta y}\,\d x\,\d y -\beta
\int\limits_{\Gamma}\left|\psi_n(x)\right|^2\e^{-2\beta y}\,\d s
$$
 % -------------- %
and using Stokes formula, $\iint\limits_{\Omega} - F_y(x,y)\,\d x\d y = \int\limits_{\partial\Omega}F(x,y) \,\d x$, we get
 % ------------ %
\[
\beta^2\iint\limits_{\Omega}
\left|\psi_n(x)\right|^2\e^{-2\beta y}\,\d x\,\d y =
\frac{\beta}{2}\int\limits_{\Gamma}
\left|\psi_n(x)\right|^2\e^{-2\beta y}\,\d x
\]
 % ------------ %
and, henceforth,
 % ------------ %
\[
S[f_n]= \iint\limits_{\Omega}
\left|\psi'_n(x)\right|^2\e^{-2\beta y}\,\d x\,\d y +\beta
\int\limits_{\R}\(\Gamma'_1(s)-1\)\left|\psi_n(\Gamma_1(s))\right|^2\e^{-2\beta\Gamma_2(s)}\,\d s\,.
\]
 % ------------ %
In the limit $n\to\infty$ the first term tends to zero and the second one to
 % ------------ %
$$
\beta\int\limits_{\R}\(\Gamma'_1(s)-1\)\e^{-2\beta\Gamma_2(s)}\,\d s\,.
$$
 % ------------ %
The integrand in this expression is non-positive, and since by assumption the trivial case, $\Gamma=\R$, is excluded, there is an interval where $\Gamma'_1(s)<1$. Consequently, $S[f_n]<0$ holds for all large enough $n$.

Under the assumption (ii) we proceed in a similar way, but we modify the trial function (\ref{f_n}) as follows,
 % ------------ %
$$
f_n(x,y)=\psi_n(x)\e^{-\gamma y}\,,
$$
 % ------------ %
where the constant $\gamma$ will be specified later. Then the form value $S[f_n]$ is
 % ------------ %
$$
\iint\limits_{\Omega}
\left|\psi'_n(x)\right|^2\e^{-2\gamma y}\,\d x\,\d y +
\int\limits_{\R}\(\frac{\beta^2+\gamma^2}{2\gamma}
\Gamma'_1(s)-\beta\)\left|\psi_n(\Gamma_1(s))\right|^2\e^{-2\gamma\Gamma_2(s)}\,\d s\,.
$$
 % ------------ %
As before, the first summand tends to zero as $n\rightarrow\infty$, while the limit of the second one is
 % ------------ %
\[
\int\limits_{\R}\(\frac{\beta^2+\gamma^2}{2\gamma}
\Gamma'_1(s)-\beta\)\e^{-2\gamma\Gamma_2(s)}\,\d s\,.
\]
 % ------------ %
In the previous case we had $\Gamma_1' = 1$ outside a compact and
we had to choose $\gamma=\beta$ to make the integral converge. Now
any $\gamma>0$ will do to the presence of the exponential factor,
recall that $\Gamma_2(s)\sim|s|\sin\alpha$ for large $|s|$.
Choosing again $\gamma = \beta$ we get as in the previous case
 % ------------ %
\[
S[f_n] \rightarrow
\beta\int\limits_{\R}\( \Gamma'_1(s)-1\)\e^{-2\beta\Gamma_2(s)}\,\d s<0
\]
 % ------------ %
as $n\to\infty$. This concludes the proof. \hfill $\Box$

\medskip

On the other hand the assumption (iii) alone does guarantee absence of a discrete spectrum. If $\Gamma$ is sufficiently smooth, we know from
Theorem~\ref{main theorem_half-plane} that a sufficient condition for its existence is a local positivity of the curvature. One may conjecture that there is no discrete spectrum for $\gamma^*\le 0$. It appears that this is indeed the case, not only asymptotically.

 % -------------- %
\begin{theorem}\label{thm: absence}
In addition to (iii), assume that $\Omega$ is concave and its boundary is $C^2$-smooth, then $\sigma_\mathrm{disc}(H_\beta)= \emptyset$.
\end{theorem}
 % -------------- %

 % -------------- %
\noindent \emph{Proof:} In view of Theorem~\ref{thm: continuous spectrum} it is sufficient to check that $H_\beta+\beta^2I\ge 0$. By assumption we have $\gamma(s)\le 0$ for any $s\in\R$, an furthermore, one can introduce the curvilinear coordinated of Lemma~\ref{lemma_existence of C_u} in the entire $\Omega$, that is, with $u$ running through the interval $(0,\infty)$ because the factor $1-u\gamma(s)$ vanishes nowhere. The quadratic form associated with $H_\beta$ is given by the expression analogous to \eqref{quadratic form for q beta} with the last term missing. For our present purpose, however, it is more suitable to regard the wave functions as elements of the space $L^2(\R\times(0,\infty), (1-u\gamma(s))\d s\d u)$, in other words to write the right-hand side of \eqref{eff-wf} as $\psi(s,u)$. The quadratic form in question can be then rewritten as
 % -------------- %
\begin{eqnarray*}
\lefteqn{q_\beta[\psi] = \int_\R \int_0^\infty \left[ \frac{1}{1-u\gamma(s)} \left| \frac{\partial\psi}{\partial s}(s,u)\right|^2 + (1-u\gamma(s)) \left| \frac{\partial\psi}{\partial u}(s,u)\right|^2 \right] \d s\,\d u} \\ && \qquad - \beta \int_\R |\psi(s,0)|^2 \d s\,. \phantom{AAAAAAAAAAAAAAAAAAAAAAAAA}
\end{eqnarray*}
 % -------------- %
Note that the expression contains the curvature but not its derivatives, hence the $C^2$ smoothness of the boundary is sufficient.  The first term on the right-hand side is obviously non-negative which allows us to estimate the form from below,
 % -------------- %
$$
q_\beta[\psi] \ge \int_\R \d s \left\{ \int_0^\infty \left| \frac{\partial\psi}{\partial u}(s,u)\right|^2 (1-u\gamma(s))\,\d u - \beta |\psi(s,0)|^2 \right\}
$$
 % -------------- %
for any $\psi\in L^2(\R\times(0,\infty), (1-u\gamma(s))\d s\d u)$. Next we notice that for a fixed $s$ the expression in the curly bracket is nothing else than the zero angular momentum part of the quadratic form associated with the Robin problem in the exterior to the disc of radius $-\gamma(s)^{-1}$. In Example~\ref{ex:disc} below we will show that the corresponding spectral threshold is larger than $-\beta^2$ the saturation being reached in the case of an infinite radius when the disc becomes a halpflane. Since $\gamma(s)\le 0$ holds for any $s\in\R$, the claim is proved.
\hfill $\Box$

%%%%%%%%%%%%%%%%%%%%%%%%%%%%%%%%%%%%%%%%%%%%%%%%%%%%%%%
\section{Exterior of a compact}\label{sect: infinite domain}

Let us turn now to the other situation mentioned in the introduction. We consider a compact a simply connected `obstacle' $\Omega$ the boundary of which is a simple close $C^4$-smooth curve $\Gamma$ and ask about the problem in its exterior, $\Omega^\mathrm{c}:= \R^2\setminus \bar\Omega$, i.e.
 % ------------ %
\begin{eqnarray}
-\Delta f=\lambda f\; &\textrm {in}& \Omega^\mathrm{c}, \nonumber \\
[-.5em] && \label{problem_ext} \\ [-.5em] \frac{\partial
f}{\partial n}=\beta f\; &\textrm{on}& \partial \Omega=\Gamma.
\nonumber
\end{eqnarray}
 % -------------- %
As before the boundary can be parametrized by its arc length, its orientation being chosen in such a way that the obstacle $\Omega$ lies to the
left as one moves along its perimeter in the positive direction. The normal $\frac{\partial}{\partial n}$ in (\ref{problem_ext}) is the outside one
\emph{with respect to} $\Omega^\mathrm{c}$. The associated quadratic form is
 % -------------- %
\begin{equation}\label{differential form_ext}
q_{\beta,\mathrm{ext}}[f] =\|\nabla
f\|^2_{L^2(\Omega^\mathrm{c})}-\beta\int\limits_{\Gamma}|f(x)|^2\d s
\end{equation}
 % -------------- %
being defined on with $\mathrm{Dom}(q_{\beta,\mathrm{ext}}) = H^{1}(\Omega^\mathrm{c}).$ It is closed and below bounded; we denote by $H_{\beta, \mathrm{ext}}$ the unique self-adjoint operator associated with this form and by $\lambda_{j,\mathrm{ext}}$ its $j$-th eigenvalue, numbered in the ascending order with the multiplicity taken into account. It is straightforward to see that $\sigma_\mathrm{ess}(H_{\beta,\mathrm{ext}})=[0,\infty)$; as before we are interested in the asymptotic behavior of the discrete spectrum as $\beta\rightarrow\infty$. To state the result, we employ again a one-dimensional comparison operator with the curvature-induced potential, this time
 % -------------- %
\begin{equation}\label{S_ext}
S=-\frac{\d^2}{\d s^2}-\frac{1}{4}\gamma^2(s) \quad
\textrm{in}\;\, L^2(0,L)\,,
\end{equation}
 % -------------- %
$L$ being the perimeter of $\Omega$, with the domain
 % -------------- %
\begin{equation}\label{P_ext}
\mathrm{Dom}(S)=\left\{f\in H^2(0,L):\ f(0)=f(L),\; f'(0)=f'(L)\right\}\,.
\end{equation}
 % -------------- %
The spectrum of $S$ is purely discrete; we denote by $\mu_j$ the $j$-th eigen\-value of $S$ counted with the multiplicity, $j\in\mathbb{N}$. The oriented curvature of $\Gamma$ of the boundary is denoted by $\gamma_\mathrm{ext}(s) := \Gamma_1'(s)\Gamma_2''(s) - \Gamma_2'(s)\Gamma_1''(s)$, and furthermore, we introduce the symbols $\gamma^*_\mathrm{ext} =\max\limits_{[0,L]}\gamma_\mathrm{ext}(s)$ and $\gamma_{*,\mathrm{ext}}=\min\limits_{[0,L]}\gamma_\mathrm{ext}(s)$.

 % -------------- %
\begin{remark}
{\rm There is no need to put the `ext' label to the operator $S$ because it is invariant with respect to the curvature sign and orientation choice. On the other hand, attention has to be paid to the curvature. We define it here in the way consistent with the convention of Sec.~\ref{sect: half-plane}, so that it is positive when the curve is turning left in the direction of the parametrization. We prefer to label it to avoid a confusion when comparing the result to that of Ref.~\cite{EMP}. In that paper the orientation of $\Gamma$ is the same, clockwise, but the curvature is defined with the opposite sign. This choice together with the opposite orientation of the normal means we have $\int_\Gamma \gamma_\mathrm{ext}(s)\,\d s=2\pi$, in particular, that the curvature is non-negative if the obstacle $\Omega$ is convex.}
\end{remark}
 % -------------- %

 % -------------- %
\begin{theorem}\label{main theorem_ext}
Under the stated assumptions, to any fixed integer $j$ there is a $\beta_j>0$ such that the number of negative eigenvalues of $H_{\beta,\mathrm{ext}}$ is not smaller than~$j$. The $j$-th eigenvalue behaves in the limit $\beta\to\infty$ as
 % -------------- %
$$
\lambda_{\mathrm{ext},j}(\beta) = -\beta^2+\gamma_{*,\mathrm{ext}}\beta+\OO\big(\beta^{2/3}\big)\,,
$$
 % -------------- %
where the lower asymptotic bound can be be improved to
 % -------------- %
$$
\lambda_{\mathrm{ext},j}(\beta) \ge
-\(\beta-\frac{\gamma_{*,\mathrm{ext}}}{2}\)^2+\mu_j+
\OO\(\frac{\log\beta}{\beta}\) \,.
$$
 % -------------- %
\end{theorem}
 % -------------- %

 % -------------- %
\begin{remark}
{\rm As in similar situations we get also an upper bound analogous
to the last formula with $\gamma_{*,\mathrm{ext}}$ replaced by
$\gamma^*_\mathrm{ext}$ which is not of much use because the two
squeeze to produce a true asymptotics only if $\Omega$ is a
circular disc. What is more important, similarly as in
Ref.~\cite{EMP} the assumption about simple connectedness of the
boundary was done for simplicity only. If $\Omega$ is a finite
family of obstacles which do not touch each other, we have the
analogous result with the asymptotics being determined by the
external curvature minimum taken over all obstacle `components'.}
\end{remark}
 % -------------- %

 % -------------- %
\noindent \emph{Proof:}
The method of Ref.~\cite{EMP}, modified in the previous sections, applies readily; we have just to change signs at appropriate places and sketch the argument briefly. We employ the map $\Phi_{\mathrm{ext}}$ defined as
 % -------------- %
\[
(0,L)\times(0,a)\ni(s,u)\mapsto(\Gamma_1(s)+u\Gamma_2'(s),\Gamma_2(s)-u\Gamma_1'(s))\in\mathbb{R}^2.
\]
 % -------------- %
It may not be defined for all $a$, of course, unless $\Omega$ is convex, but due to the smoothness of the boundary it is a diffeomorphism for $a$ small enough. We choose such an $a>0$ to be sufficiently small and denote by $\Sigma_{a,\mathrm{ext}}$ the one-sided strip neighborhood of $\Gamma$ of width
$a$,
 % -------------- %
\[
\Sigma_{a,\mathrm{ext}}:=\Phi_{\mathrm{ext}}((0,L)\times(0,a))\,,
\]
 % -------------- %
and as before we impose Dirichlet and Neumann conditions at the curve $\Gamma_a :=\Phi_{\mathrm{ext}}((0,L)\times\{a\})$ which is at the same time the boundary of the unbounded and simply connected domain $\Omega^\mathrm{c}\setminus\overline\Sigma_{a,\mathrm{ext}}=:\Lambda_{a,\mathrm{ext}}$. Denoting by $q_{a,\beta,\mathrm{ext}}^{D/N}[f]$ the corresponding quadratic forms supported on $\Sigma_{a,\mathrm{ext}}$, and by $L_{a,\beta,\mathrm{ext}}^{D/N}$, respectively, the associated self-adjoint operators, we have
 % -------------- %
\begin{equation}\label{inequalities H_beta_ext}
L_{a,\beta,\mathrm{ext}}^{N}\oplus(-\Delta^N_{\Lambda_{a,\mathrm{ext}}})\leq
H_{\beta,\mathrm{ext}}\leq L_{a,\beta,\mathrm{ext}}^{D}\oplus
(-\Delta^D_{\Lambda_{a,\mathrm{ext}}})
\end{equation}
 % -------------- %
in $L^2(\Omega^\mathrm{c})= L^2(\Sigma_{a,\mathrm{ext}})\oplus L^2(\Lambda_{a,\mathrm{ext}})$, where the parts related to $\Lambda_{a,\mathrm{ext}}$ are positive and can be thus neglected. Next we pass to the curvilinear coordinated using the formula analogous to \eqref{eff-wf} with the opposite sign in the denominator and rewrite the quadratic form supported by the strip neighborhood. For any $f\in H^{1}(\Sigma_{a,\mathrm{ext}})$ we also have
$\varphi\in H^{1}((0,L)\times(0,a))$ and the form equals
 % -------------- %
\begin{eqnarray}\nonumber
\lefteqn{\int\limits_{0}^L\int\limits_0^a \left[
\frac{1}{(1+u\gamma)^2}\left|\frac{\partial\varphi}{\partial
s}\right|^2 +
\left|\frac{\partial\varphi}{\partial u}\right|^2 +  V_{\mathrm{ext}}
\left|\varphi\right|^2 \right](s,u)\, \d u\,\d s} \nonumber \\ &&
-\int\limits_{0}^L\(\beta-\frac{\gamma(s)}{2}\)|\varphi(s,0)|^2\d
s  -\int\limits_{0}^L\frac{\gamma(s)}{2(1+a\gamma(s))}|\varphi(s,a)|^2\d
s\,, \phantom{AAA} \label{quadratic form for q beta_ext}
\end{eqnarray}
 % -------------- %
where
 % -------------- %
\[
V_{\mathrm{ext}}(s,u)=-\dsfrac{\gamma^2(s)}{4(1+u\gamma(s))^2}+\frac{u\gamma''(s)}{2(1+u\gamma(s))^3}
-\frac{5}{4}\frac{u^2(\gamma'(s))^2}{(1+u\gamma(s))^4}\,.
\]
 % -------------- %
In analogy with Lemma~\ref{l: uniteq} we introduce operators in $L^2((0,L)\times(0,a))$ unitarily equivalent to $L_{a,\beta,\mathrm{ext}}^{D/N}$ and the quadratic forms associated with them. To pass to estimating operators with separated variables, we introduce
 % -------------- %
\begin{eqnarray*}
&& \gamma_+:=\max\limits_{[0,L]}|\gamma(.)|\,, \quad \gamma'_+:=\max\limits_{[0,L]}|\gamma'(.)|\,,
\quad \gamma''_+:=\max\limits_{[0,L]}|\gamma''(.)|\,, \\
&& V_{\mathrm{ext},+}(s):=-\frac{\gamma^2(s)}{4(1+a\gamma_+)^2}+\frac{a\gamma''_+}{2(1-a\gamma_+)^3}\,, \\
&&
V_{\mathrm{ext},-}(s):=-\frac{\gamma^2(s)}{4(1-a\gamma_+)^2}-\frac{a\gamma''_+}{2(1-a\gamma_+)^3}
-\frac{5}{4}\frac{a^2(\gamma'_+)^2}{(1-a\gamma_+)^4}\,.
\end{eqnarray*}
 % -------------- %
We choose an $a$ satisfying $0<a<\gamma_+/2$ and for vectors $\varphi$ belonging to the domains
$Q_a^D$ and $Q_a^N$ defined in analogy with \eqref{Qformdomains}, respectively, we define
 % -------------- %
\begin{eqnarray*}
\lefteqn{\widetilde b_{a,\beta,\mathrm{ext}}^D[\varphi] =
(1-a\gamma_+)^{-2}\iint\limits_{0\ 0}^{\ \ \ a\ L}
\left|\frac{\partial\varphi}{\partial s}\right|^2\d s\,\d u+
\iint\limits_{0\ 0}^{\ \ \ a\ L}
\left|\frac{\partial\varphi}{\partial u}\right|^2\d s\,\d u} \\ &&
+ \iint\limits_{0\ 0}^{\ \ \ a\ L}V_{\mathrm{ext},+}(s)
\left|\varphi\right|^2\d s\,\d u-\(\beta -
\frac{\gamma^*}{2}\)\int\limits_0^L|\varphi(s,0)|^2\,\d s
\end{eqnarray*}
 % -------------- %
and
 % -------------- %
\begin{eqnarray*}
\lefteqn{\widetilde b_{a,\beta,\mathrm{ext}}^N[\varphi]=
(1+a\gamma_+)^{-2}\iint\limits_{0\ 0}^{\ \ \ a\ L}
\left|\frac{\partial\varphi}{\partial s}\right|^2\d s\,\d u+
\iint\limits_{0\ 0}^{\ \ \ a\ L}
\left|\frac{\partial\varphi}{\partial u}\right|^2\d s\,\d u} \\ &&
+ \iint\limits_{0\ 0}^{\ \ \ a\ L}V_{\mathrm{ext},-}(s)
\left|\varphi\right|^2\d s\,\d
u -\(\beta-\frac{\gamma_*}{2}\)\int\limits_0^L|\varphi(s,0)|^2\,\d
s \\ && \quad -\frac{\gamma_+}{2(1-a\gamma_+)}\int\limits_0^L|\varphi(s,a)|^2\,\d
s\,.
\end{eqnarray*}
 % -------------- %
The forms $\widetilde b_{a,\beta,\mathrm{ext}}^D[\varphi]$ and $\widetilde b_{a,\beta,\mathrm{ext}}^N[\varphi]$ are similar to those denoted by the same symbols in Ref.~\cite{EMP}, the only difference is that $\gamma^*$ is replaced with $-\gamma_*$ and $\gamma_*$ with $-\gamma^*$, this following the argument of the said paper we get, in particular, the lower bound stated in Theorem~\ref{main theorem_ext}.

To get a better upper bound which allows to get the tow term asymptotic expansion we employ again a variational estimate choosing trial functions of the form
 % -------------- %
$$
\hat\varphi(s,u):=\chi\(\dsfrac{s-s^*+(2j-1)\varepsilon}{2\varepsilon}\)\(\e^{-\alpha
u}-\e^{-2a\alpha+u\alpha}\)\,,
$$
 % -------------- %
where $\chi$ is again a fixed smooth function with the support in $(0,1)$ and $s_*$ is the point in which the curvature reaches its minimum,
$\gamma(s_*)=\gamma_*$. For $j=1$ we get
 % -------------- %
\begin{eqnarray*}
\lefteqn{\dsfrac{b_{a,\beta,\mathrm{ext}}^D[\hat\varphi]}{\|\hat\varphi\|^2_{L^2(0,L)}}\le
\dsfrac{1}{4\varepsilon^2}\
\dsfrac{\|\chi'\|_{L^2(0,1)}^2}{\|\chi\|_{L^2(0,1)}^2}\(1+\OO\big(\alpha^{-1}\big)+C\varepsilon\(\dsfrac{1}{2\alpha}+\mathrm{\alpha^{-2}}\)\)}
\\ && +\alpha^2\(1+\OO\big(\alpha\e^{-2a\alpha}\big)\) +
\(-\dsfrac{\(\gamma_*\)^2}{4}+C\varepsilon\)\(1+\OO\big(\alpha^{-1}\big)\)
\\ &&
-2\alpha\(\beta-\dsfrac{\gamma_*+\varepsilon}{2}\)\(1+\OO\big(\alpha\e^{-2a\alpha}\big)\)
%\label{ineq 1}
\end{eqnarray*}
 % -------------- %
and choosing $\alpha=\beta+\dsfrac{\gamma_*}{2}$ the right-hand side of the estimates becomes
 % -------------- %
\begin{eqnarray*}
\lefteqn{\hspace{-2em} \dsfrac{1}{4\varepsilon^2}\
\dsfrac{\|\chi'\|_{L^2(0,1)}^2}{\|\chi\|_{L^2(0,1)}^2}\(1+\OO\big(\beta^{-1}\big)+C\varepsilon\(\dsfrac{1}{2\beta}+\mathrm{\beta^{-2}}\)\)
-\(\beta-\dsfrac{\gamma_*}{2}\)^2} \\ &&
+\varepsilon\(\beta-\dsfrac{\gamma_*}{2}\)
+\(\dsfrac{-\(\gamma_*\)^2}{4}+C\varepsilon\)\(1+\OO\big(\alpha^{-1}\big)\)\,.
\end{eqnarray*}
 % -------------- %
Optimizing with respect to $\varepsilon$ by taking $\varepsilon=\beta^{-1/3},$ we get the inequality
 % -------------- %
\begin{equation} \label{finalbound_}
\dsfrac{b_{a,\beta}^D[\hat\varphi]}{\|\hat\varphi\|^2_{L^2(0,L)}}\le-\(\beta-\dsfrac{\gamma_*}{2}\)^2+\OO\big(\beta^{2/3}\big)
\end{equation}
 % -------------- %
which the result for the first eigenvalue. The argument for $j\ge 2$ is the same, as before we take into account that by construction the used trial functions are mutually orthogonal. \hfill $\Box$

\medskip

We have mentioned that the Dirichler-Neumann estimates used in the proof squeeze only in the case when the curvature is constant. Let us look at this situation more closely.

 % -------------- %
\begin{example} \label{ex:disc}
{\rm Suppose that the obstacle $\Omega$ is a disc of radius $R$, for definiteness centered at the origin, hence
 % -------------- %
\begin{equation}\label{gamma explicit}
\gamma(s)\equiv\gamma=\frac{1}{R}
\end{equation}
 % -------------- %
and the the comparison operator $S$ is just a shifted Laplacian with periodic boundary conditions and its eigenvalues $\mu_j$ can be expressed  explicitly as
 % -------------- %
\begin{equation}\label{mu_j explicit}
\mu_j=\(-\ \frac{1}{4}+\left[\frac{j}{2}\right]^2\)R^{-2},
\end{equation}
 % -------------- %
where $[y]$ denotes the maximum integer which less or equal to $y$. The rotational symmetry makes it natural to employ polar coordinates,
 % -------------- %
\[
\left\{\begin{array}{ccc}x=r\cos
\theta\\y=r\sin\theta\end{array}\right. \qquad r\geq R\,,\
0\leq\theta<2\pi\,.
\]
 % -------------- %
Writing with an abuse of notation $f(x,y)=v f(r,\theta)$ we can cast the eigenvalue problem in question with $\lambda=-k^2$ into the form
 % -------------- %
\begin{equation}\label{Laplacian polar coordinates and boundary condition}
\left\{\begin{array}{l}\dsfrac{\partial^2f}{\partial
r^2}+\frac{1}{r}\frac{\partial f}{\partial r}+ \frac{1}{r^2}\
\frac{\partial^2f}{\partial\theta^2}=k^2f,\\\\\left.-\dsfrac{\partial
f}{\partial r}\right|_{\ r=R}=\beta f.\end{array}\right.
\end{equation}
 % -------------- %
Solution to the first equation in (\ref{Laplacian polar coordinates and boundary condition}) is conventionally sought in the form $f(r,\theta)= \sum_{m\in\mathbb{Z}}c_m K_m(kr)\e^{\i m\theta}$. Moreover, operator commutes with the angular momentum, $-i \frac{\partial}{\partial\theta}$ with periodic boundary conditions, hence they have common eigenspaces, and we can consider sequence $\{c_m\}$ with nonzero $c_m$ corresponding to a single values of $|m|$; it goes without saying that the discrete spectrum has multiplicity two except the eigenvalue corresponding to $m=0$ which is simple. The boundary condition in (\ref{Laplacian polar coordinates and boundary condition}) can be then rewritten as
 % -------------- %
\begin{equation}\label{equation I_m (1)}
k K'_m(kR)+\beta K_m(kR)=0.
\end{equation}
 % -------------- %
for a fixed $m\in\Z$. To find its solutions, let us change the variables to $u=kR$ and $\alpha=\beta R,$ in which case the condition
(\ref{equation I_m (1)}) reads
 % -------------- %
\begin{equation}\label{equation I_m (2)}
-\frac{u K'_m(u)}{K_m(u)}=\alpha.
\end{equation}
 % -------------- %
The function at the left-hand side is strictly increasing for $u>0$, equal to $m$ at $u = 0$, hence (\ref{equation I_m (2)}) has a unique solution for any fixed $m$ and $\alpha>m$. As $\alpha\rightarrow+\infty,$ so does $u$ in (\ref{equation I_m (2)}), and using the well-known asymptotics of modified Bessel functions, we find
 % -------------- %
$$
-\frac{uK'_m(u)}{K_m(u)}=u+\frac{1}{2}+\frac{4m^2-1}{8u}+O(u^{-2})\,,\quad
u\rightarrow+\infty\,.
$$
 % -------------- %
In combination with the spectral condition (\ref{equation I_m (2)}) this yields
 % -------------- %
$$
u=\alpha-\frac{1}{2}-\frac{4m^2-1}{8\alpha}+O(\alpha^{-2})\,,\quad\alpha\rightarrow+\infty\,.
$$
 % -------------- %
This, in turn, implies the asymptotics for $u^2$, and returning to the original variables $\beta, k$ we find
 % -------------- %
$$
-k^2=
-\(\beta-\frac{1}{2R}\)^2+\(m^2-\frac{1}{4}\)R^{-2}+\OO(\beta^{-1})\,,\quad\beta\rightarrow+\infty\,.
$$
 % -------------- %
This agrees, of course, with the conclusion of Theorem~\ref{main theorem_ext} taking (\ref{gamma explicit}) and (\ref{mu_j explicit}) into account. At the time it shows that there is not much room for improving the error term in the theorem, because it differs from the one in this explicitly solvable example by the logarithmic factor only.

The result also shows that eigenvalues corresponding to the exterior of a disc are larger that $-\beta^2$. This is true not only asymptotically, which is the fact we have used in the proof of Theorem~\ref{thm: absence}. Indeed, from (\ref{equation I_m (2)}) we can derive that $0<u<\alpha$, or equivalently, $0<k<\beta$ and $-\beta^2 < -k^2$. It follows from the fact that
 % -------------- %
\begin{equation}\label{toprove}
-\frac{K_m'(x)}{K_m(x)}>1,\quad m\in\mathbb{N}\,, x>0\,.
\end{equation}
 % -------------- %
To check this inequality we use the relations \cite[p.79, (26),
p.82, (21)]{BE}
 % -------------- %
$$
K_{m-1}(x) + K_{m+1}(x) = -2
K'_{m}(x)\,,\quad K_{m}(x)=\int\limits_{0}^{\infty} \e^{-x \cosh u}
\cosh (mu) \d u\,.
$$
 % -------------- %
Using the first of these formul{\ae} we can rewrite (\ref{toprove}) as $K_{m+1}(x)-K_{m}(x)>K_{m}(x)-K_{m-1}(x).$ The validity of the last relation, in turn, follows from the positivity of the second derivative of $K_m(x)$ with respect to $m$, which can be checked using the second one the above
formul{\ae}.

}
\end{example}

\subsection*{Acknowledgments}
The research has been supported by  the projects
``Support of inter-sectoral mobility and quality enhancement of
research teams at Czech Technical University in Prague'',
CZ.1.07/2.3.00/30.0034, sponsored by European Social Fund in the
Czech Republic, and \mbox{14-06818S} of the Czech Science
Foundation. A.M. expresses his gratitude to Stepan Manko and
Satoshi Ohya for useful discussions.

\end{document}